\documentclass[article,useAMS,usenatbib,usegraphicx]{mn2e}

\usepackage{slashbox}
\usepackage{longtable}

\title[Rotation in NGC 2264]{Rotation in NGC 2264: a study based on CoRoT\thanks{The CoRoT space mission, launched on 2006 December 27, was developed and is operated by the CNES, with participation of the Science Programs of ESA, ESA's RSSD, Austria, Belgium, Brazil, Germany and Spain} photometric observations}
\author[L. Affer, G. Micela, F. Favata, E. Flaccomio]{L. Affer$^{1}$\thanks{E-mail:
affer@astropa.inaf.it},  G. Micela$^{1}$, F. Favata$^{2}$, E. Flaccomio$^{1}$, J. Bouvier$^{3}$\\
$^{1}$Istituto Nazionale di Astrofisica, Osservatorio Astronomico di Palermo G.\,S. Vaiana, Piazza del Parlamento 1, 90134 Palermo, Italy\\
$^{2}$European Space Agency, 8-10 rue Mario Nikis, 75738 Paris Cedex 15, France\\
$^{3}$UJF-Grenoble 1/CNRS-INSU, Institut de Plan\`etologie et d'Astrophysique de Grenoble (IPAG) UMR 5274, Grenoble, F-38041, France}
\begin{document}

\date{Accepted 2013 January 2 - Received 2012 December 20; in original form 2012 July 31}

\pagerange{\pageref{firstpage}--\pageref{lastpage}} \pubyear{2002}

\maketitle

\label{firstpage}

\begin{abstract}
Rotation is one of the key stellar parameters which undergo substantial evolution during the stellar lifetime, in particular during the early stages.  
Stellar rotational periods can be determined on the basis of the periodic modulation of starlight produced by non-uniformities on the surface of the stars, due to manifestation of stellar activity. We present the results of an extensive search for rotational periods among NGC 2264 cluster members, based on photometric monitoring using the CoRoT satellite, with a particular 
attention to the distribution of classical and weak-line T-Tauri stars. NGC 2264 is one of the nearest and best studied star forming region in the solar neighbourhood, with an estimated age of 3 Myr, and is the object of a recent simultaneous multiband campaign 
including a new CoRoT observation with the aim to assess the physical origin of the observed variability. We find that the rotational distributions of classical and weak-line T-Tauri star are different, suggesting a difference in the rotational properties of accreting and non-accreting stars.\\

\end{abstract}

\begin{keywords}
clusters: NGC 2264 -- stars: rotation -- stars: pre-main sequence.
\end{keywords}

\section{Introduction}\label{intro}

In the last years, the determination of stellar rotation rates for large samples of stars with different masses and ages in young open clusters, has made substantial progress, providing a large observational sampling of the angular momentum evolution of pre-main sequence 
stars (\citealt{bou08, bou09}). The precise mechanisms governing angular momentum evolution of pre-main sequence low-mass stars are still not well understood, but basically they can be schematized with two main competing processes: star contraction and star-disk interaction. 
Stars in the pre-main sequence phase are still contracting, thus increasing their angular velocity to conserve the angular momentum. 
On the other hand, the observed existence of slow rotators and of a wide dispersion in rotation rates of cluster stars on the zero-age main sequence (ZAMS), can be explained only with the presence of a competing mechanism of angular momentum loss (i.e. spin down of the star), 
different from star to star, during the pre-main sequence phase. It is generally believed that this mechanism can be explained by the magnetic interaction between stellar magnetospheres
and circumstellar disks, in a scenario known as {\it disk-locking}, first proposed by \citet{cam90} and \citet{kon91} and explained in detail by
\citet{sno+94}, which assumes that the angular momentum deposited on an accreting star (due to mass accretion from disk to star,
\citealt{esh+93}) is exactly removed by torques carried along magnetic field lines connecting the star to the disk.
The wide dispersion of rotational velocities observed on the ZAMS is the result of different disk lifetimes (\citealt{bcf+93}; \citealt{ccq+95}). 
Several observational results indicate a relation between the presence of disks and rotational evolution, in particular population of stars with
disks, on average, rotate more slowly than those without disks and does exist a statistically significant anti-correlation between angular velocity and disk indicators such as
near-infrared excess and ${H{\alpha}}$ equivalent width (\citealt{esh+93}; \citealt{bcf+93}; \citealt{hrh+00}; \citealt{hbm+02}; \citealt{lmb+05}; \citealt{rsm+06}). Moreover,
the disk-locking scenario predicts that the torques arising from the magnetic connection between the star and the disk remove substantial angular
momentum enforcing an equilibrium angular spin rate (\citealt{c&h96}) which is in agreement with the constant rotation period in the 2-8 days range,
characteristic of the majority of young stars. However, there have been several conflicting theoretical and observational evidences concerning the role of
disk-locking scenario in the evolution of low mass pre-main sequence stars.
\citet{ds05} pointed out that the P$_{rot}$ distribution histograms for weak T-Tauri stars (WTTSs, whose periodic variability is believed to be induced by large starspots)
and classical T-Tauri stars (CTTSs, whose variability may be also due to accretion spots and shadowing of the photosphere from dusty disk structures) in NGC 2264 are very similar and do not 
indicate that CTTSs are rotating more slowly than their WTTS counterparts. Furthermore,  \citet{ds05} did not find a correlation between P$_{rot}$ and
theoretical age, as might be expected if stars were spinning up after decoupling from their disks. \citet{smm+99} and \citet{c&b06} did not find a
correlation between accretion and rotation in ONC and IC 348 low mass stars, respectively (though they do not conclude that their results are inconsistent
with disk-locking). We have to note
that most of the samples for which there is no clear evidence of a connection between the existence of disks and slow rotation, suffer from several
biases, such as small sample size, sample biased toward small-mass or high-mass stars, the use of NIR photometry as disk indicator or the use of $v sin
i$ values instead of rotation periods, which make them unsuited to perform this kind of test, on star-disk interaction outcomes. In particular, the presence of a near-infrared
excess does not garantee that the star is actually accreting mass from a disk.
The studies of \citet{c&b07} and \citet{rsm+05}, based on {\it Spitzer} mid-infrared observations, however, as well as demonstrating that objects which currently show mid-infrared excesses
are more likely accreting than not, also found differences in the rotational properties of accreting and non accreting stars for NGC
2264 and the Orion Nebula Cluster (ONC), respectively, and represent the best test case to date, providing the strongest evidence that star-disk interaction regulates the angular
momentum evolution of pre-main sequence stars. \\ 
The idea and the basic assumptions of disk-locking, sketched above,
are a simplification of a much more complex phenomenon, and indeed several discussions on the shortcomings of the theory and its confrontation with
observations have been put forward. In particular, \citet{mpd+10,mpg+12} critically examined the theory of disk locking, noting that the 
differential rotation between the star and disk naturally leads to an opening (i.e., disconnecting) of the magnetic field between the two. 
They find that this significantly reduces the spin-down torque on the star by the disk, thus, disk-locking cannot account (at least, alone) for the slow rotation 
observed in several systems and for which the model was originally developed. \citet{mpd+10,mpg+12} supported the idea that stellar winds may be
important during the accretion phase, they may be powered by the accretion process itself and be the key driver of angular momentum loss (\citealt{h&m82};
\citealt{p&c96}; \citealt{m&p05}). A strong magnetically driven wind, as proposed by \citet{m&p05}, is an idea which deserves further study, as well as
the development of a more realistic theorical model able to explain the full range of observed rotation periods and magnetic phenomena and the achievement of a sufficient amount of accurate data to empirically constrain
them. \\
NGC 2264 is one of the best known studied star forming regions in the solar vicinity (d$\approx$\, 760 pc, age $\approx$\, 3 Myr) and is considered a benchmark for the study of star formation processes in our Galaxy. 
NGC 2264 luckily falls in the small portion of the sky accessible by CoRoT (COnvection ROtation and planetary Transits, \citealt{baglin06}), and thus
represents a unique chance for the mission and the study of young stars still in a formation phase.
Its distance and age make it an ideal CoRoT target, its size is well suited to the CoRoT field of view, with a large fraction of cluster members falling in the appropriate magnitude range for accurate photometric monitoring in the CoRoT observations.\\ 
NGC 2264 has been extensively observed at all wavelengths from radio to X-rays (see \citealt{dah08}, for a review on the region), for studies of the star formation process through the observation of its outcomes: the Initial Mass Function (e.g. \citealt{sbc+08}), 
the star formation history, and the spatial structure (e.g. \citealt{tly+06}; \citealt{ssb09}). NGC 2264 is also a primary target for the study of the evolution of the stellar angular momentum and its relation to circumstellar accretion (e.g. \citealt{lmb+05}), of the evolution (and dispersal) 
of circumstellar disks (e.g. \citealt{atg+10}) and of the correlation between optical and X-ray variability of young stars (\citealt{fmf+10}).
Among several optical, IR, and X-ray surveys, both photometric and spectroscopic, on NGC 2264, the most recent are: \citet{sbc+08}, who has provided the widest area and deepest publicy available optical photometry; \citet{ssb09}, who has published {\it Spitzer} (IRAC + MIPS) photometry; \citet{rms+02} 
and \citet{ds05}, who have published spectral types, H$\alpha$, and Li equivalent widths, from low-dispersion spectra, for a total of $\sim$ 500 members; \citet{fhl+06}, who has published radial velocities for 436 stars.\\
\citet{lbm+04} performed a photometric monitoring of about 10600 stars to search for
periodic and irregular variable pre-main sequence stars and found 543 periodic variables with periods between 0.2 days and 15 days, and 484 irregular variables. \citet{lmb+05} used this extensive study to conclude that the period distribution in NGC 2264 is similar to that of
the Orion Nebula Cluster, though shifted toward shorter periods (confirming the conclusion of \citealt{sks+99}, based on the analysis of the rotation rates of 35 candidate members, that the stars in NGC 2264 are spun up with respect to members of the Orion Nebula Cluster).\\ 
The period distribution found by \citet{lmb+05} is unimodal for masses lower than 0.25 M$_{\odot}$\, while it is bimodal for more massive stars. \citet{lmb+05} also found evidence for disk locking with a constant period, among 30\% of the higher mass stars (with a locking period of $\sim$8 days), 
while disk-locking is less important among lower mass stars, whose peak in the period distribution at 2-3 days, suggests that these stars have undergone a low rate of angular momentum loss from star-disk interaction, while not completely locked (an evolution scenario defined by \citet{lmb+05} as ``moderate'' angular momentum loss).\\ 
The bimodality is interpreted as an effect of disk-star interactions in pre-main sequence stars, slow rotators being interpreted as stars that are magnetically locked to their disks, preventing them from spinning up with time and accounting for the broad period distribution
of ZAMS stars. This assumption is supported by some observational results showing that WTTSs are rotating faster than CTTSs with inner disks (\citealt{esh+93}; \citealt{bcf+93}). Nevertheless, the hypothesis that accreting stars rotate more slowly than non accreting
ones is still a matter of debate, since conflicting evidence exists (see e.g. \citealt{lmb+05}; \citealt{ds05}; \citealt{rsm+05,rsm+06};
\citealt{smm+99}; \citealt{c&b06,c&b07}), as explained above.\\
The CoRoT satellite has allowed us to conduct a large scale survey of photometric variability of NGC 2264. Thanks to the accurate high-cadence photometry and large field of view of CoRoT, we could study rotation and activity of about 8000 stars in a 3.4 sq.degree region. The entire star-forming region fits into a single CoRoT field of view, 
and the campaign resulted in continuous 23-day light curves for 301 known cluster members brighter than V$\approx$\,16 (M=0.3-0.4 M$_{\odot}$). The resulting optical broad-band light curves are the first accurate, highest cadence (32 or 512 seconds), longest duration, data set available for $\approx$\,3 Myr old stars.\\ So far, the CoRoT NGC 2264 
data have been used to study the correlation between optical and X-ray variability in young stars (\citealt{fmf+10}); asteroseismological properties of two high mass cluster members (\citealt{zkg+11}); and to identify and study the behaviour of NGC 2264 members that are AA Tau-like (\citealt{atg+10}).\\ \citet{atg+10} demonstrated that the peculiar 
photometric behaviour of AA Tau, which consist in a flat maximum in the light curve interrupted by deep quasi-periodical minima (due to obscuring material with a variable structure which is located in the inner disk region, near the corotation radius), that vary in depth and width from one rotational cycle to the other, is quite common among CTTSs (\citealt{bca+99, bga+03}; 
\citealt{mbd+03}; \citealt{bab+07}; \citealt{gbm+07}). The interpretation for AA Tau can now be considered quite solid, and it's extension to an high fraction of CTTSs simply requires that the size of the obscuring clump of material (e.g. the height of the inner disk warp) is larger than previously thought. \\ 
In a rather similar scenario of circumstellar material orbiting the star and consequent time-variable shading, \citet{fmf+10} found
evidence of a correlation between soft X-ray and optical variability of CTTSs (no correlation is apparent in the hard band), while no
correlation in either band (soft and hard) is present in WTTSs. \citet{fmf+10} suggested that this observation is consistent with a
scenario in which a significant fraction of the X-ray and optical emission from CTTSs is affected by temporally variable shading and
obscuration.\\ The conclusions of both \citet{atg+10} and \citet{fmf+10} point toward a different origin of the observed periods, suggesting a difference both in the physical and morphological properties of CTTSs and WTTSs.\\
In this paper we derive accurate rotational periods of known NGC 2264 members, testing
whether a relationship between accretion and rotation exists in pre-main sequence stars.\\
In the following we describe the observational and data reduction strategy (Sect.~\ref{obs} and \ref{red}). In Section~\ref{discuss} we discuss the results obtained, and our major conclusions are summarized in Sect.~\ref{concl}. 

\begin{figure}
   \centering
   \includegraphics[width=8cm]{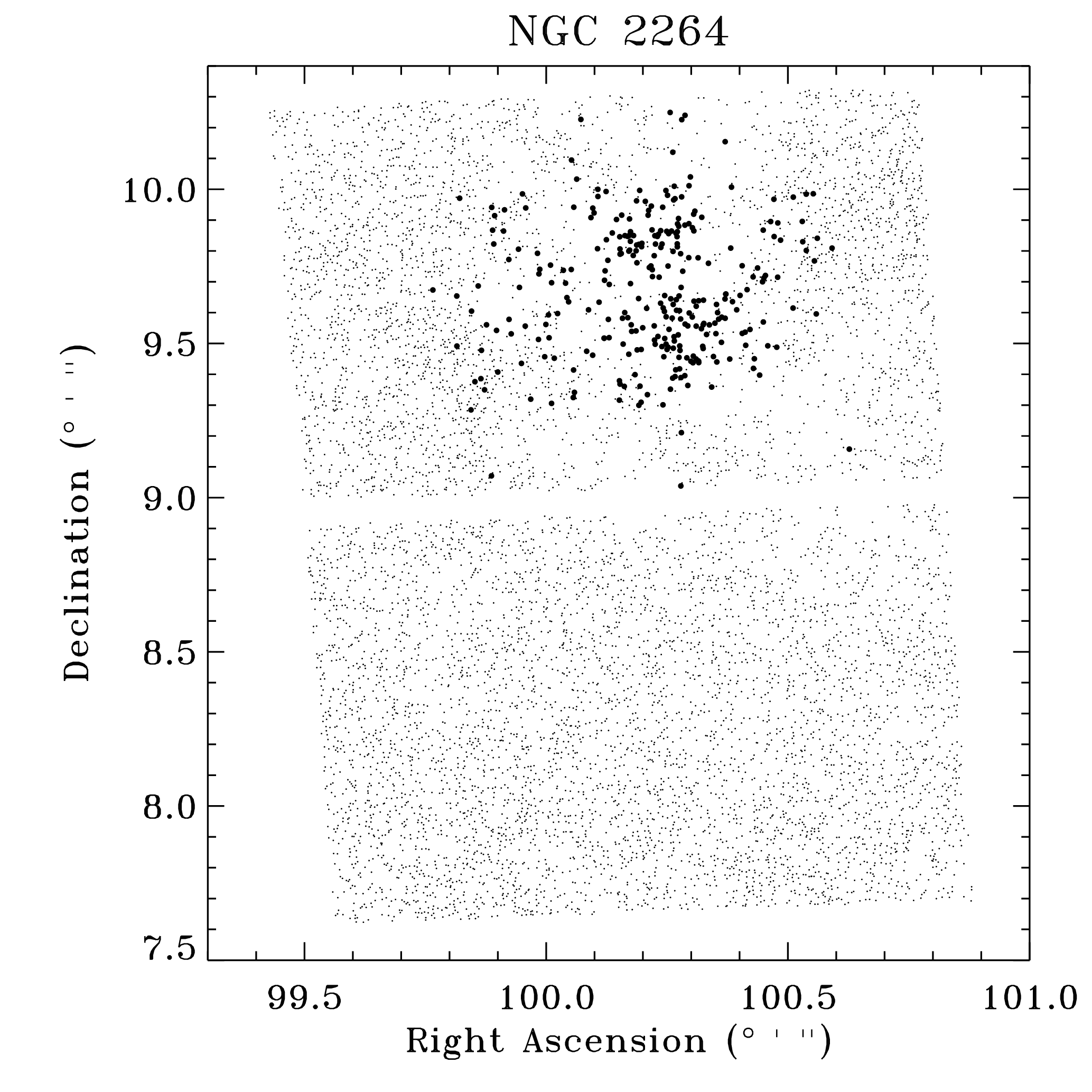}
      \caption{Spatial distribution of the CoRoT targets in the SRa01 (dots) with target stars satisfying one or more membership criteria (big dots), as described in the text.
              }
         \label{fig1}
   \end{figure}

\section[]{CoRoT observations}\label{obs}
The first short run CoRoT observations (SRa01, P.I. F.Favata) lasted from March 7 to March 31, 2008 (23 days) and was devoted to the observation of the very young ($\sim$\, 3 Myr) stellar open cluster NGC 2264, which covers most of the mass sequence from $\sim$\, 3 to
$\sim$\, 0.1 M$_{\odot}$. \\
A total of 8150 stars were observed, with right ascension (RA) between 99.4${^\circ}$ and 100.9${^\circ}$ and
declination (DEC) between 7.6${^\circ}$ and 10.3${^\circ}$ and R magnitudes from 9.2
to 16.0. The sample includes 301 cluster members (see Sec.~\ref{red} for details regarding membership criteria). We used the so called N2 data delivered by the CoRoT pipeline (\citealt{sfc+07}) after correction of the electronic offset, gain, electromagnetic interference, and outliers. The pipeline includes background subtraction and partial jitter
correction. Low quality data points, e.g. taken during the South Atlantic Anomaly crossing or due to hot pixels events, are flagged. Some of the stars have light curves in three separate but ill-defined bands (red, green, and blue). In these cases our analysis was conducted on the white-light data obtained by summing the three bands. The light curves are sampled at a rate of 512 s or oversampled at 32 s. All the light curves presented here were rebinned to 512 sec.
Using CoRoT photometry we are able to reveal luminosity
variation, with a precision down to 0.1 mmag per hour (magnitude between 11 and 16), during continuous observations, allowing to measure photometric periods also in relatively quiet stars (for comparison, the luminosity
variations of the Sun range  between  $\sim$0.3 mmag and $\sim$0.07 mmag at maximum and minimum activity, respectively, \citealt{aig04}).

\begin{figure*}
   \centering
   \includegraphics[width=12cm]{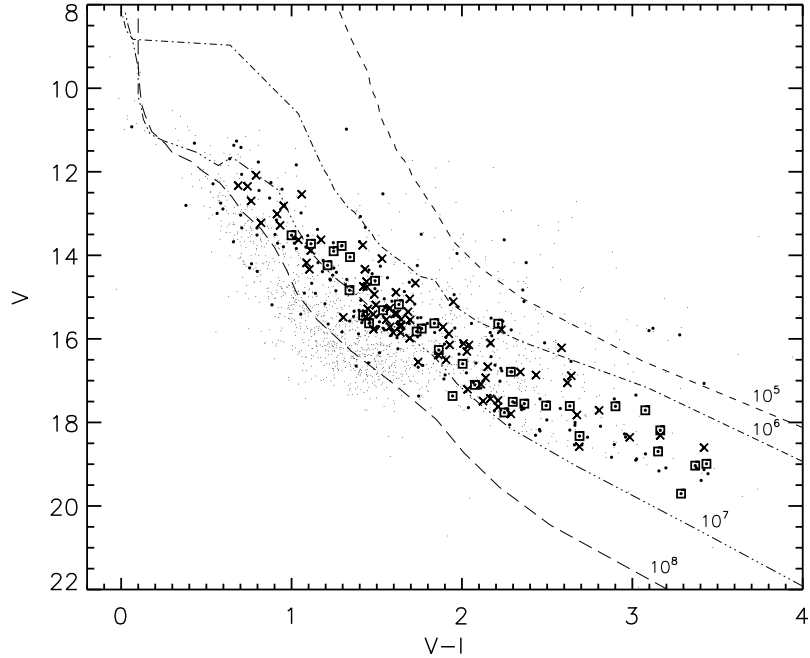}
      \caption{The colour-magnitude diagram of the 8150 SRa01 stars observed by CoRoT (gray dots). The black large dots indicate the 301 cluster stars satisfying one or more of the membership criteria. Crosses are WTTSs and squares are CTTSs, defined following our criterium (WTTSs for EW$_{H{\alpha}}$ $\le$\,5 \AA\, and CTTSs for EW$_{H{\alpha}}$ $\ge$\,10 \AA). The solid lines denotes isochrones (yr) from \citet{sfd97} (transformed to the observational plane using the \citealt{kh95} compilation).
              }
         \label{fig2}
   \end{figure*}

\section[]{Sample extraction and light curve analysis}\label{red}
We selected 301 cluster stars (whose spatial distribution is shown in Fig.~\ref{fig1} as big dots) satisfying
one or more of the following membership criteria (the number of objects which fulfill the various criteria is indicated near each criterium):

\begin{itemize}
\item Detection in X-rays by Chandra ACIS or XMM-Newton (\citealt{rrs+04}; \citealt{fms06} + Flaccomio et al., in preparation; \citealt{dsp+07}) and
location on the cluster sequence in the (I, R-I) diagram, when I and R magnitudes are available (191);
\item High levels of H$\alpha$\, emission, indicative of accretion, according to photometric indices (\citealt{lbm+04}; \citealt{sbc+08}) (104);
\item H$\alpha$\, with spectroscopic equivalent width greater than 10 \AA\, and/or indicated to be in strong emission by \citet{fhl+06} (86); 
\item Classified as Class I or Class II according to \citet{ssb09}, based on Spitzer mIR photometry (76);
\item Strong optical variability + high H$\alpha$\, emission, indicative of high chromospheric activity, according to \citet{lbm+04} (87);
\item Radial velocity members according to \citet{fhl+06} (192).
\end{itemize}
After selecting the cluster members, we classified them as CTTSs if their H$\alpha$\, equivalent width was greater than 10 \AA, and WTTSs
if smaller than 5 \AA. The threshold between these two classes is a function of spectral type, as suggested by \citet{mar98} and deeply analyzed by
\citet{b&m03}. The information regarding the H$\alpha$\, equivalent width are available from the work of \citet{ds05} for 164
members, 86 with EW$_{H{\alpha}}$ $\le$\,5 \AA, 19 with EW$_{H{\alpha}}$ between 5 and 10 \AA\, and 59 with EW$_{H{\alpha}}$ $\ge$\,10 \AA.
We decided to exclude intermediate EWs  (5 $<$\, EW$_{H{\alpha}}$ $<$\, 10 \AA), though not a usual procedure when dealing with H$\alpha$, to keep the two sample of CTTSs and WTTSs well separate and thus avoid possible ambiguity in the classification.
In Fig.~\ref{fig2} we show the V vs. (V-I) colour-magnitude diagram for the 8150 stars in the SRa01 observations with isochrones from \citet{sfd97} (transformed to the observational plane using the \citealt{kh95} compilation), together with the NGC 2264 members. \\
We have analyzed the CoRoT light curves (LCs) as in \citet{amf+12}, we refer the reader to this work for a full
description of data reduction and analysis. In brief, we prepared the light curves by correcting the following systematic effects: spurious data points (mainly due
to cosmic rays and/or to the satellite crossing of the South Atlantic anomaly), as flagged by the reduction pipeline were removed; we rebinned the data to two
hours to smooth out the orbital period of the satellite (1.7 h); spurious long period trends (due to pointing and instrumental drift) were removed by fitting a third degree polynomial to the data and then dividing the original data points by this function.\\

\begin{figure*}
   \centering
   \includegraphics[width=7.cm]{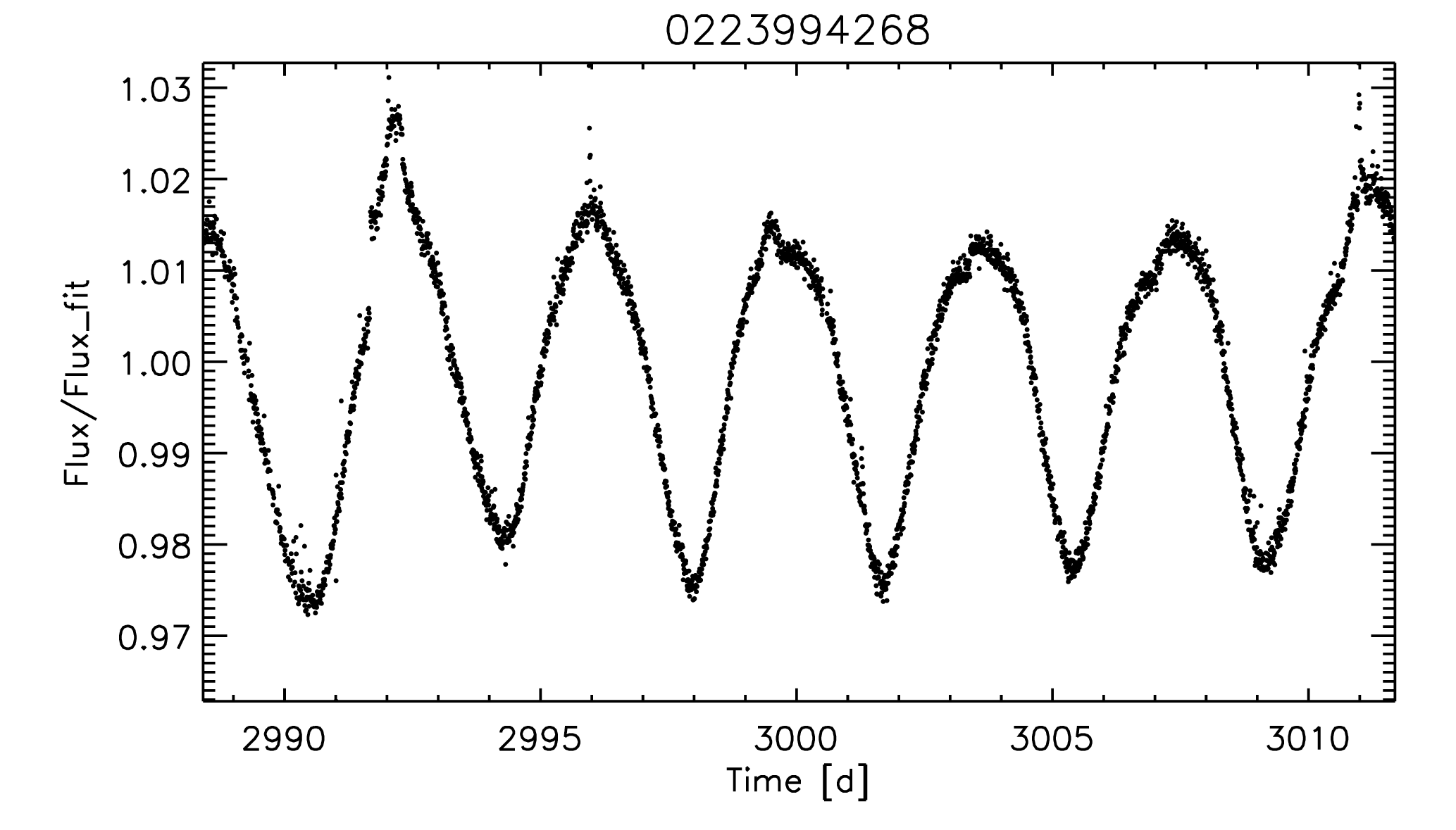}
\hspace{1.0cm}
   \includegraphics[width=7.cm]{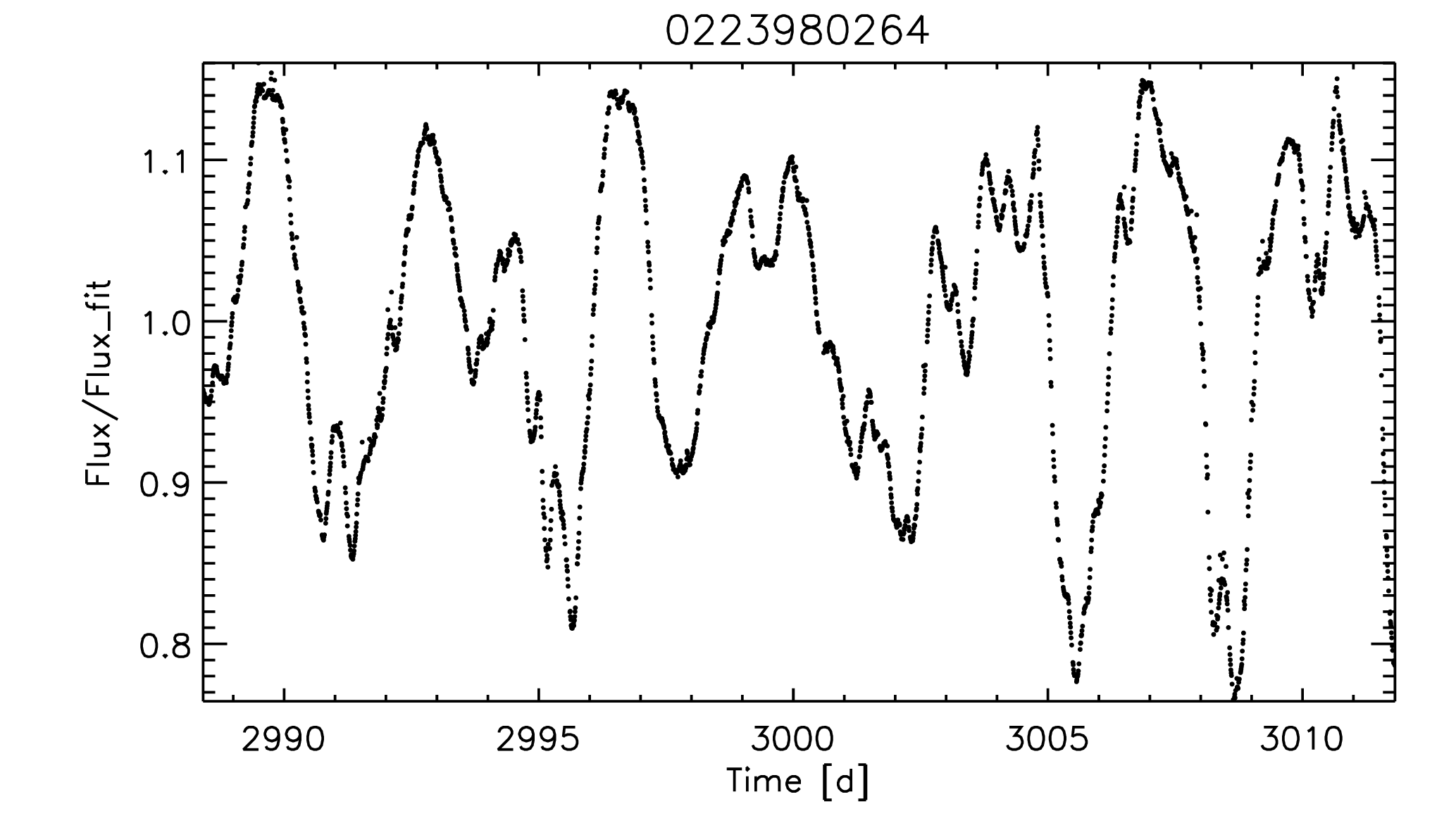}
\vspace{0.5cm}
   \includegraphics[width=7.cm]{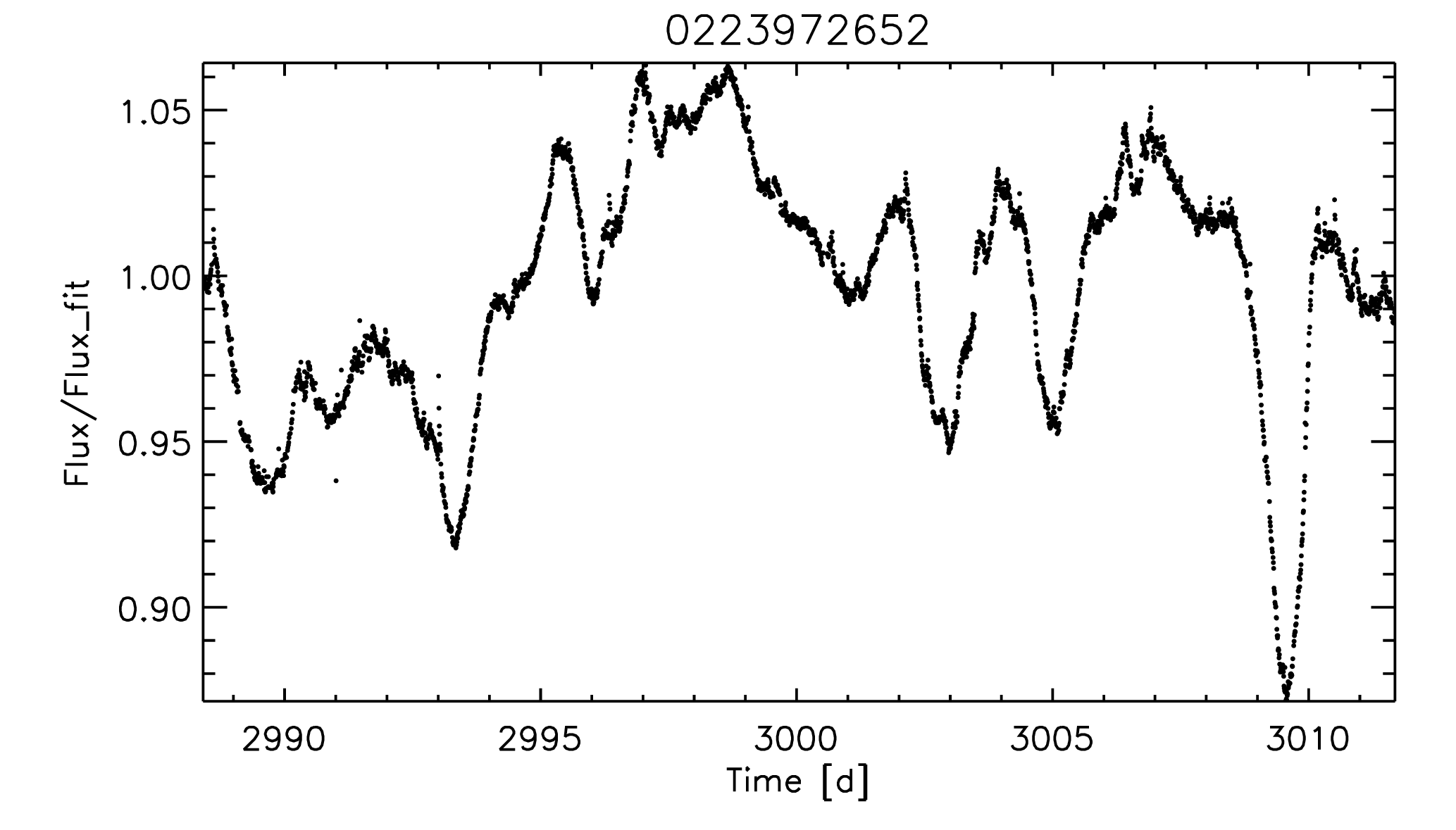}
\hspace{1.0cm}
   \includegraphics[width=7.cm]{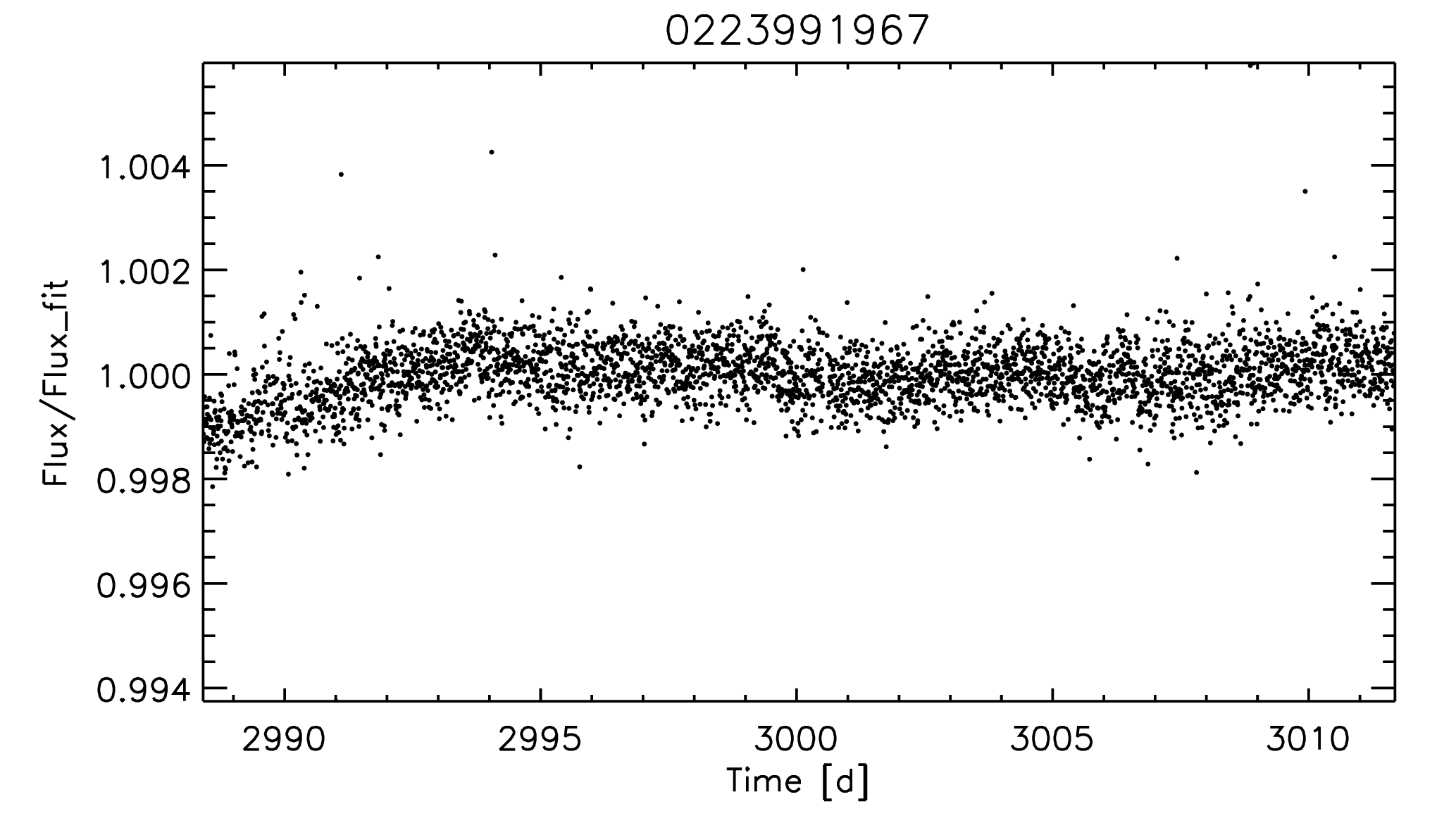}
      \caption{Four examples of morphologically different LCs (rebinned to 512 s): spot-like periodic (top left
      panel); AA Tau-like system (top right panel); irregular (bottom left panel) and non-variable
      ``noisy-like'' (bottom right panel). Time is in days from 2000 January 1 (JD-2451544.5).
      }
         \label{fig3}
   \end{figure*}

The presence of periodic signals was detected using the Lomb Normalized Periodogram (LNP) technique 
(\citealt{sca82}, \citealt{h&b86}). With this algorithm we calculated the normalized power $P(\omega)$ as function of
angular frequency $\omega\,= 2\pi\,\nu$\, and identified the location of the highest peak in the
periodogram. In order to decide the significance of the peak we have followed  \citet{ehh95}, 
randomizing the temporal bins from the original light curve. By calculating the maximum power on a large number of randomized data sets, the conversion
from power to False Alarm Probability (FAP) can be determined. In detail,
we constructed 1000 light curves \emph{resampled} from the original ones randomizing the position of blocks of adjacent 
temporal bins (block length, 12 h) (e.g. \citealt{flaccomio05}). By shuffling the data we break any possible time correlation and periodicity of the light curve on time scales longer
than the block duration. We calculated the Scargle periodogram for all the randomized light curves and
we compared the maximum from the real periodogram to the distribution obtained from the 
randomized light curves, at the same frequency, in order to establish the probability that values as high as the observed 
one are due to random fluctuations. Given some threshold FAP$_*$\, we state that the detected candidate 
periodicity is statistically not significant if FAP $>$\, FAP$_*$. The calculation we performed on CoRoT light curves led often to small FAPs, 
indicating that LNPs of our light curves present peaks that in most cases cannot be explained by pure stochastic noise, or non-periodic variability on time scales shorter than 12 h. 
In fact, if light curves present variations on time scale smaller than the size of the temporal block we used in the simulations, these variations will be 
still present in the simulated curves and will be not recognized as significant (more details in \citealt{amf+12}). We
have chosen a bin size of 12 h as a reasonable compromise between the expected time scale of stochastic variations and the shortest expected
periodic signal.
Once significant peaks (at FAP 1\%; since we simulated 1000 light curves, the 1\% FAP power is the power that was exceeded by the highest peak
in 10 simulations) were determined from the LNPs, we also used an autocorrelation analysis \citep{b&j76} to validate the periodicity of the LCs and to eliminate or correct spurious periods due to aliasing effects (in 5\% of the cases, with a 95\% confidence interval) or residual effects due to the choice of temporal blocks used in the LCs' simulations (block lenght, 12 h). \\
Autocorrelation takes each point of the light curve measured at time $t$ and compares the value of that point
to another at time $t+L$. Points separated by $L$ will have very similar values if the data contained some variability with period
$L$, thus the autocorrelation function will have peaks corresponding to periods of variability in the data. The
autocorrelation $r_{L}$ of a sample population X as a
function of the lag $L$ is:

$$ r_{L} =\frac{\sum_{k=0}^{N-L-1}(x_{k}-\bar{x})(x_{k+L}-\bar{x})}{\sum_{k=0}^{N-1}(x_{k}-\bar{x})^2}$$
where $\bar{x}$ is the mean of the sample population X and N is the sample size, the
quantity $r_{L}$ is called the autocorrelation coefficient at lag $L$. The correlogram for a time series 
is a plot of the autocorrelation coefficients $r_L$ as a function of L. A time series
is random if it consists of a series of independent observations with the same distribution. In this case we would expect the
$r_L$ to be statistically not significant for all values of $L$. 
We have chosen to adopt a 95\%  confidence level to select significant autocorrelation coefficient.\\

\begin{figure*}
   \centering
   \includegraphics[width=8cm]{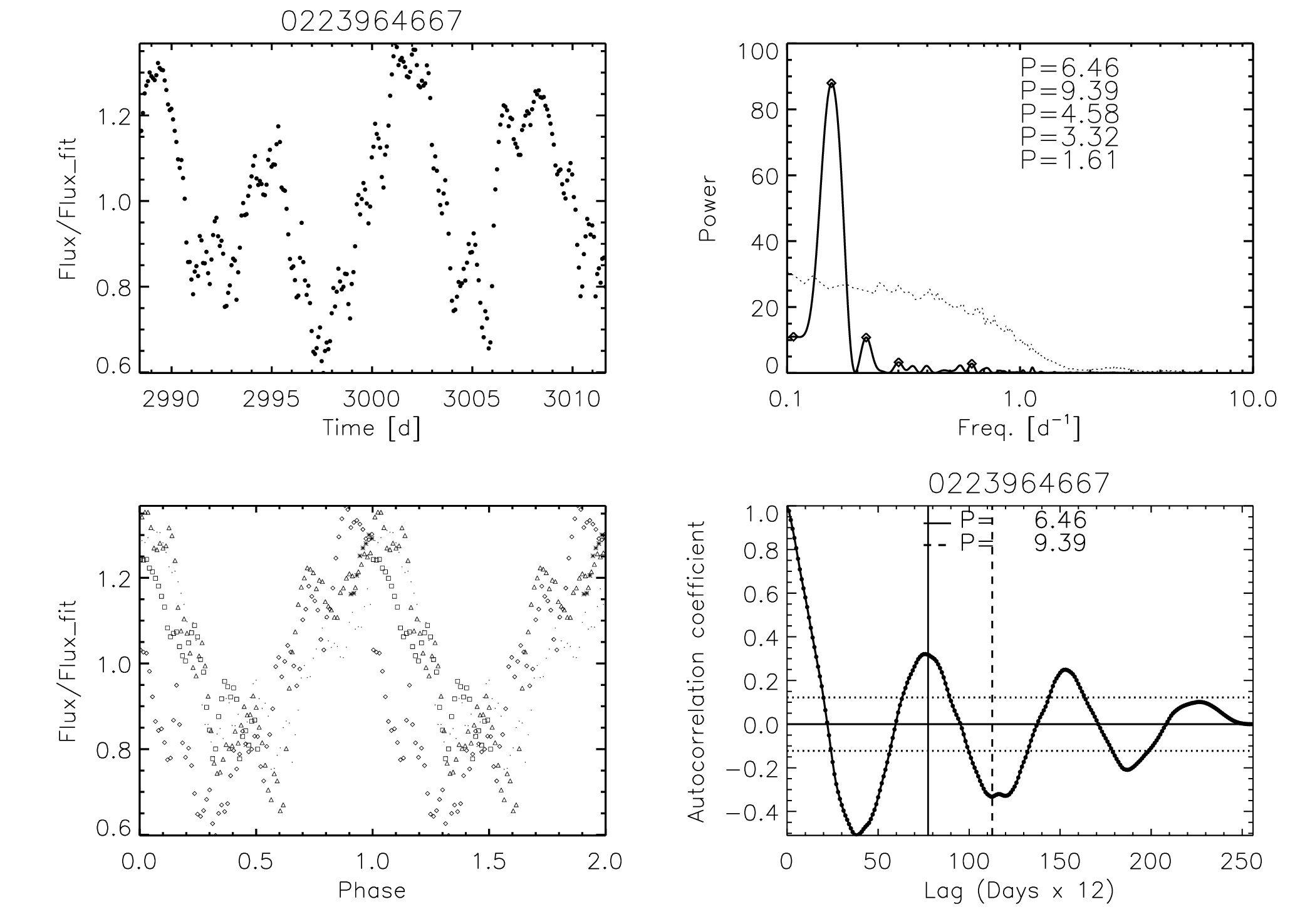}
   \includegraphics[width=8cm]{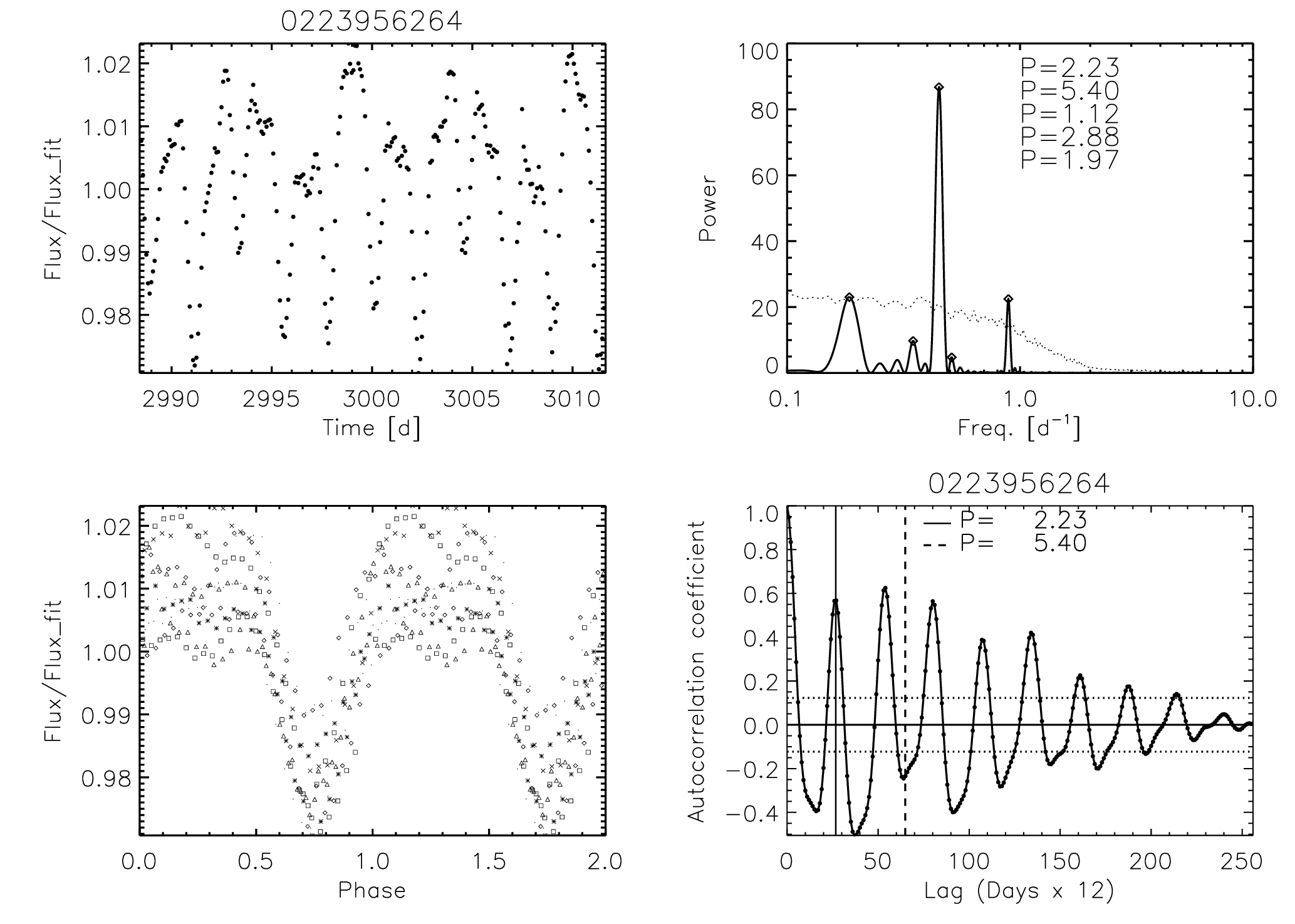}
      \caption{Two examples of NGC 2264 members: the four left plots are for a classical T-Tauri star, CoRoT ID 0223964667, while the four right
      plots are for a weak line T-Tauri star, CoRoT ID 0223956264. For each example, the top panels show the LC and the relative LNP, with the dotted curve superimposed indicating
      the 1\% significance threshold determined by simulations (the periods indicated in the top right of the LNP plot are the five most significant
      periods yielded by the LS periodogram), while the bottom panels show the LC folded with the most significant period for each example, and the autocorrelation plot
      with the 95\% confidence interval (dotted horizontal lines) with the vertical lines indicating the position of the two most significant periods
      found with the LNP. Time in the LCs is in days from 2000 January 1 (JD-2451544.5).
              }
         \label{fig4}
   \end{figure*}

\begin{figure*}
\centering
\includegraphics[width=7.5cm]{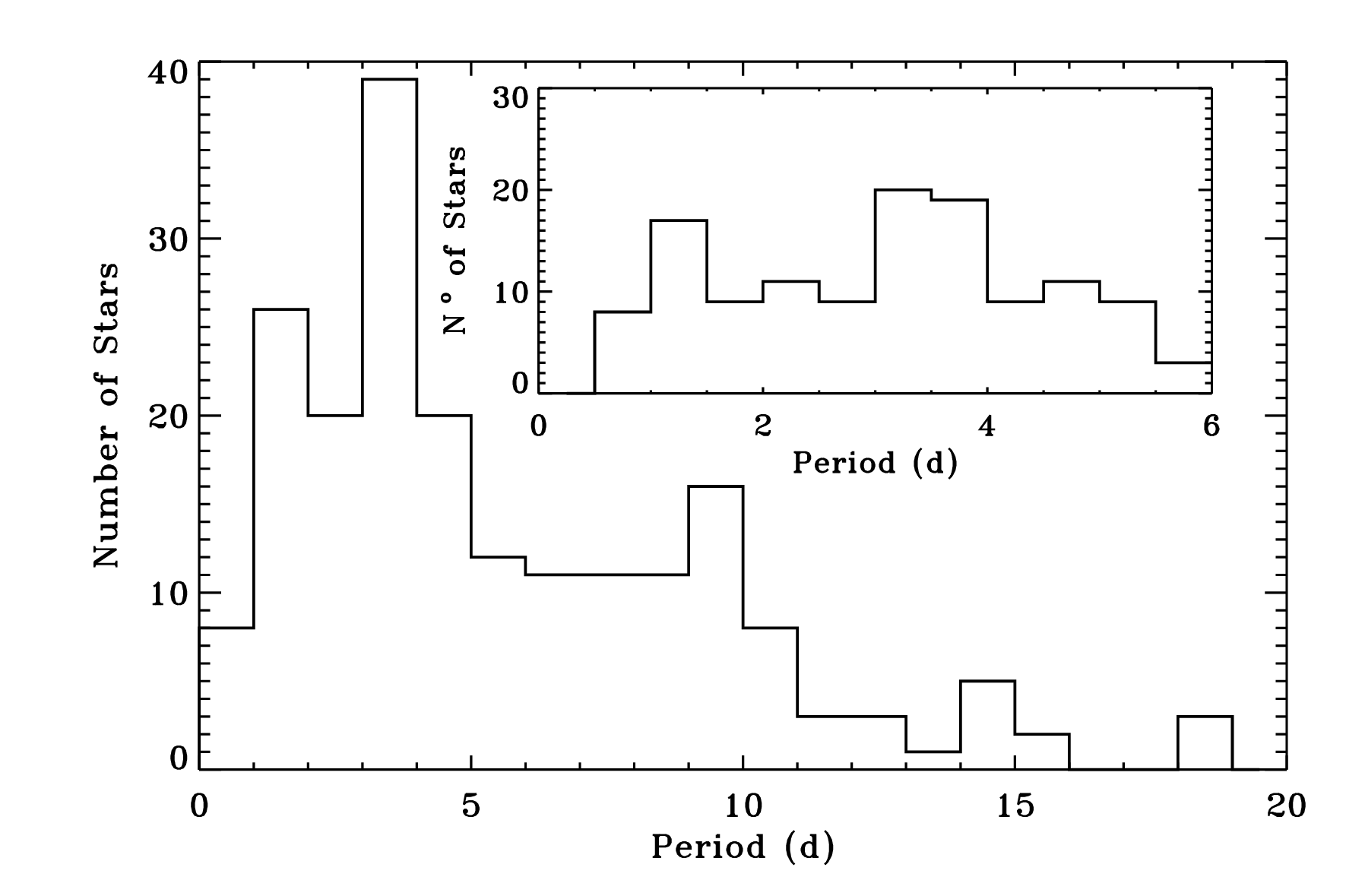}
\hspace{1.0cm}
\includegraphics[width=7.5cm]{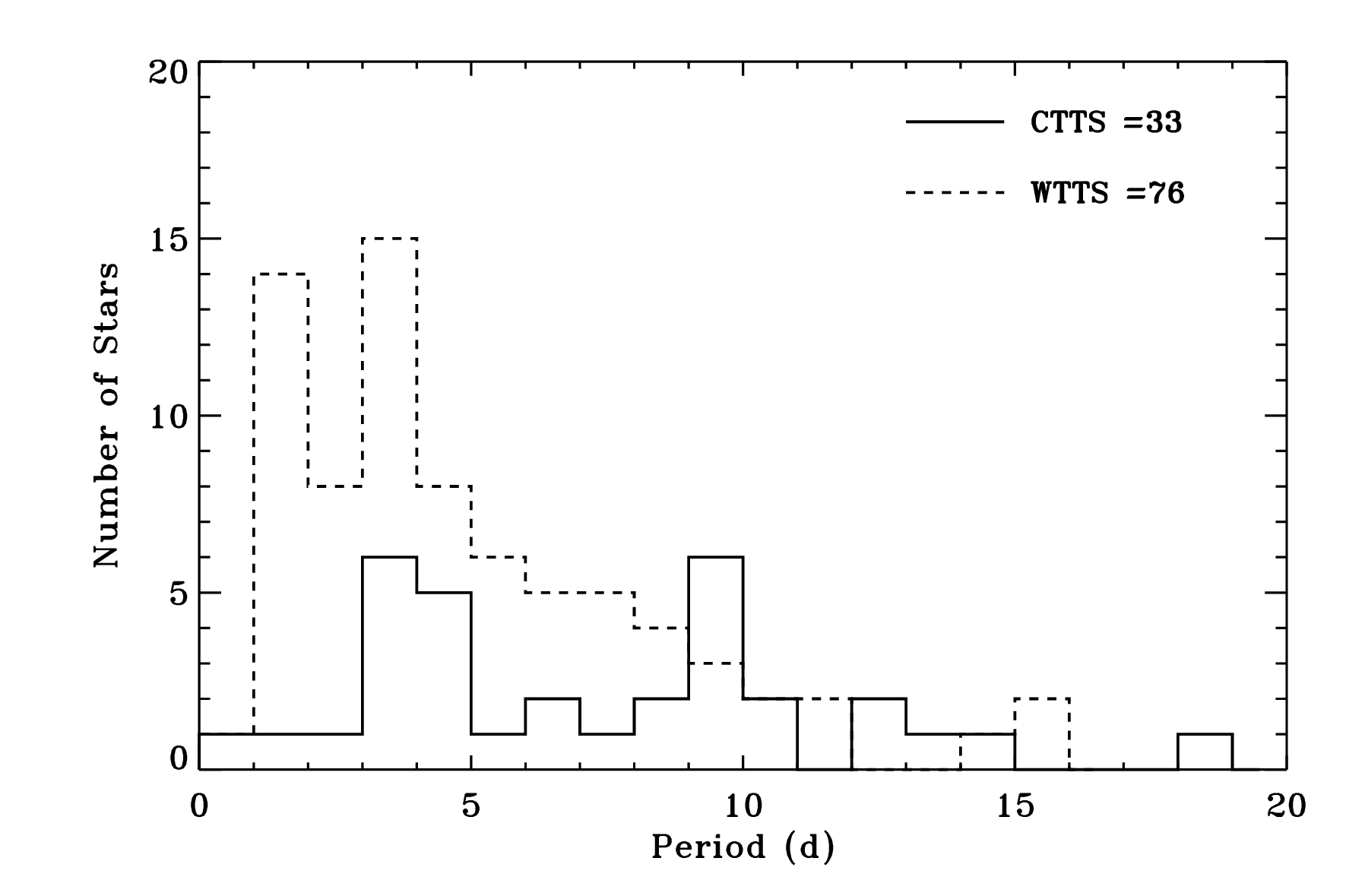}
\caption{Left panel: Rotational period distribution for NGC 2264 members, the inset plot is the same diagram but zoomed into the period region between 0-6 days. Right panel: Rotational period distributions for CTTSs (solid line) and for WTTSs (dashed line).}
 \label{fig5}
\end{figure*}

Following \citet{atg+10}, morphologically we can divide the LCs in four groups: 1) spot-like periodic LCs, whose periodicity can be interpreted as rotational modulation of surface features such as cool and/or hot spots (Fig.~\ref{fig3}, top left panel); 2) AA Tau-like systems, whose quasi-periodicity is likely caused by the obscuration of the stellar photosphere by circumstellar disk material (Fig.~\ref{fig3}, top right panel), the stability of the spot-like LCs on the timescale of the observations, makes them easily distinguishable from the AA Tau-like ones; 3) irregular LCs, whose non-periodic brightness modulations (some of them with peak-to-peak variations up to 1 mag) are probably due to a complex mixing of non-steady accretion phenomena, and obscuration by non-uniformly distributed circumstellar material, as suggested by \citet{atg+10} (Fig.~\ref{fig3}, bottom left panel); 4) non-variable LCs which do not display an obvious periodicity, most of
which looks like noisy LCs with small variability amplitude $\le$ 1\% (Fig.~\ref{fig3}, bottom right panel).\\ 
To evaluate a short term variability amplitude of the LC, we 
calculated the running median flux, obtained using a temporal bin set up by 15 time points (thus, a time scale of 15$\times$512 s), we calculated for each time
value the difference between the instantaneous flux and the running median flux derived at the corrisponding time, obtaining an array. We measured the amplitude variation of the LC as the
difference between the maximum and the minimum values of this array. We found that the variability amplitude range
is 2\% to 171\% for CTTSs and 1.0\% to 38\% for WTTSs in our sample. In Fig.~\ref{fig4} we show the results of the complete analysis for two LCs, one belonging to a CTTS and the other to a WTTS.\\ 

\begin{table*}
\centering
\caption{Samples used according to membership criteria, $H\alpha$ equivalent widths available from \citet{ds05} and membership-IR classification (Class
II and Class III) following \citet{ssb09}.}
\label{tab00}
\begin{tabular}{l|cccc|cc} \hline
Samples	 &     Spot-like&  AA-Tau like& Irregular& Non-periodic&CII& CIII\\\hline
301 members &    189  &  20  &  45 &   47 &  42&  114 \\\hline
164/301 &   &   &  &   &  &    \\
({\tiny with $H\alpha$}) &    &   &   &    &   &\\\hline
59 CTTS &    23  &    12  &  23 &   1 &  22&  6 \\
({\tiny EW $\ge$ 10 \AA}) &	   &	 &    &     &	&\\
19 intermediate &    12   &   1  &  5 &   1 &  7&  2 \\
({\tiny 5 $<$ EW $<$ 10 \AA}) &	   &   &    &	&   &\\
86 WTTS &    77   &  2  &  6 &   1 &  4&  59 \\
({\tiny EW $\le$ 5 \AA}) &    &    &	&    &   &\\\hline
\end{tabular}
\begin{flushleft}
\end{flushleft}
\end{table*}

\section{Results: Rotational Periods}\label{discuss}
Of the 301 monitored cluster members, we found that 189 are periodic variables, with regular light curves, possibly resulting from the rotational
modulation of the light by stellar spots, 20 are AA Tau-like systems, 45 are irregular LCs, and 47 display no significant variability
(noise dominated LCs). Among the 86 WTTSs, we measured periods for 76 stars (74 regular and 2 AA Tau-like), 6 were found to be irregular and 1 non-variable. For
3 stars we did measure a period, the
variability is clearly due to spot, nonetheless we discarded these periods as they are not significant according both the LNP and autocorrelation
methods. Among the 59 CTTSs, we measured
periods for 33 stars (21 regular and 12 AA Tau-like), 23 were found to be irregular and 1 non-variable, for two stars the periods found
were discarded as not significant. In Table~\ref{tab00} we list the samples used in this work, indicating the morphological division performed
(spot-like, AA-Tau like, etc.) and the membership-IR classification (Class II and Class III) following \citet{ssb09}. \\ 
All the rotational periods derived for NGC 2264 members are reported in Table~\ref{tab01}. For each star we list: the CoRoT ID; the derived period; the
variability amplitude ${AmpVar}$; the R magnitude; the B-V magnitude; the right
ascension and declination; the membership-IR classification (Class II and Class III) following \citet{ssb09} and other membership criteria listed in \citet{sbc+08}; the $H\alpha$ equivalent widths from \citet{ds05} and the mass from the \citet{sdf00} tracks.\\
In the left panel of Fig.~\ref{fig5} we show the rotational period distribution for NGC 2264 members, while the distribution of rotational periods for
CTTSs and WTTSs is shown in the right panel. Although the statistics are limited, it is evident that the period distribution of the CTTSs looks quite different from that of the WTTSs, with CTTSs being
slower rotators on average (median P$_{Rot}$ = 7.0 days) with respect to WTTSs  (median P$_{Rot}$ = 4.2 days), in the cluster NGC
2264. According to a Kolmogorov-Smirnov (K-S)
test (\citealt{ptv+02}), there is only a probability of 2\% that the two distribution are equivalent.

\begin{figure}
   \centering
   \includegraphics[width=7cm]{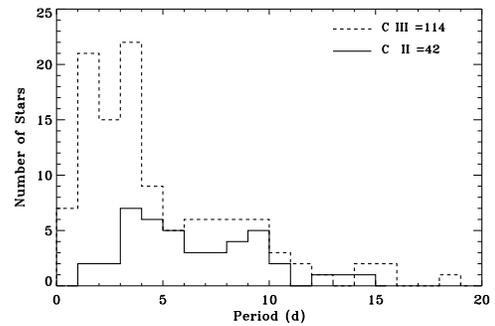}
      \caption{Rotational period distribution for CII (dashed line) and CIII members (solid line). The two class were selected following
      the membership-IR classification of \citet{ssb09} and other membership listed in \citet{sbc+08}.
              }
         \label{fig6}
   \end{figure}

\begin{figure}
   \centering
   \includegraphics[width=7cm]{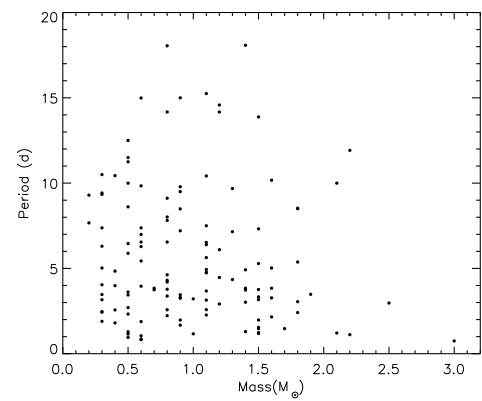}
      \caption{Rotation period as a function of stellar mass for our sample. The mass values are derived from the \citet{sdf00} tracks.
              }
         \label{fig7}
   \end{figure}

The peaks of the distribution for CTTSs are located at about $P=3-5$ days and  $P=9$ days. WTTSs distribution suggests two peaks at $P=3-5$ days and 
$P=1-2$ days. The difference is even more evident if we compare the period distributions of Class II and Class III stars, as is shown in
Fig.~\ref{fig6}. We selected Class II and Class III members following
the membership-IR classification of \citet{ssb09}, using the four $\it Spitzer$\, IRAC bands and other memberbership criteria listed in \citet{sbc+08}. 
For Class II-Class III members the K-S test yields a probability of 0.2\% that the two distributions are drawn from the same parent population.
The observed rotational period distribution are in agreement with the conclusion
derived by the ground-based study of \citet{lmb+05}, on the accretion-rotation relationship, that is, differences do exist in the rotational behaviour of
accreting and non accreting stars. \\Fig.~\ref{fig7} shows rotation period plotted
as a function of stellar mass, derived from the \citet{sdf00} tracks. The diagram shows no clear trend of rotation period with stellar
mass. We calculated the period distributions in three mass intervals: M $<$ 0.75 (``low''); 0.75 $<$ M
$<$ 1.55 (``mid''); 1.55 $<$ M $<$ 3.0 M$_{\odot}$ (``high''). Using a K-S test, we have verified that the probabilities that the ``low'' and
``mid'', the ``low'' and ``high'', and the ``mid'' and ``high'' distributions are equivalent are 91\%, 58\% and 54\%, respectively, neither of which 
can lead us to infer that they are significantly different. The median periods are 4.8, 4.5 and 3.5 days for the ``low'', ``mid'' and ``high'' samples, respectively.

\begin{figure}
   \centering
   \includegraphics[width=7cm]{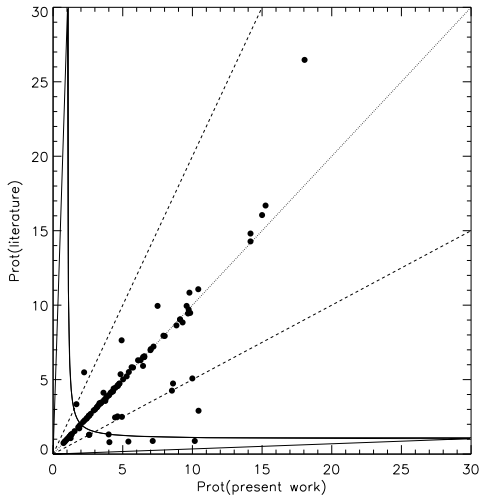}
      \caption{Comparison between the periods derived in the present study and the ones derived by \citet{lbm+04} for the sample stars in common. Stars with the same period in the two studies are located on the bisector (dotted line). The two dashed lines on each side of the bisector represent harmonics (0.5P - 2.0P). The solid curves show the loci of the 1 day$^{-1}$ aliases.
              }
         \label{fig8}
   \end{figure}

\begin{figure*}
   \centering
   \includegraphics[width=9cm,height=2.5cm]{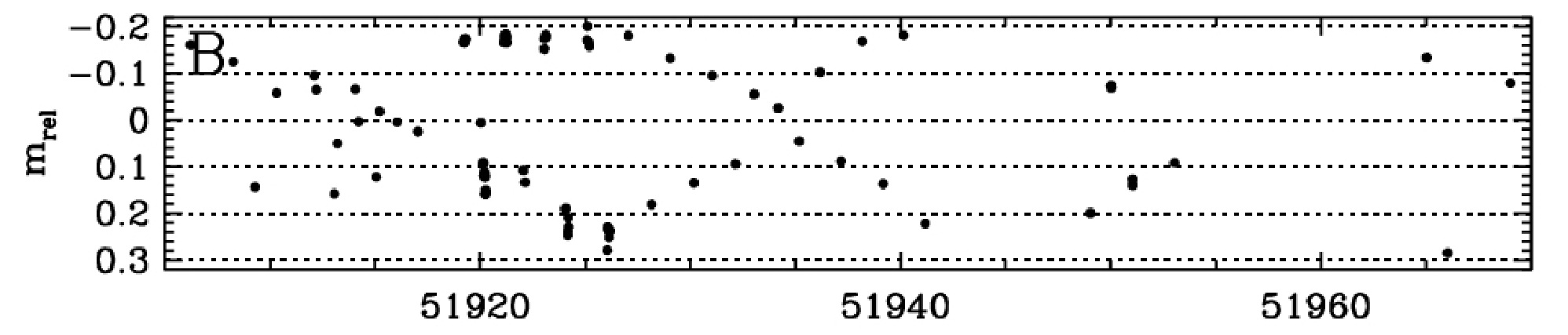}
   \hspace{0.2cm}
   \includegraphics[width=5cm]{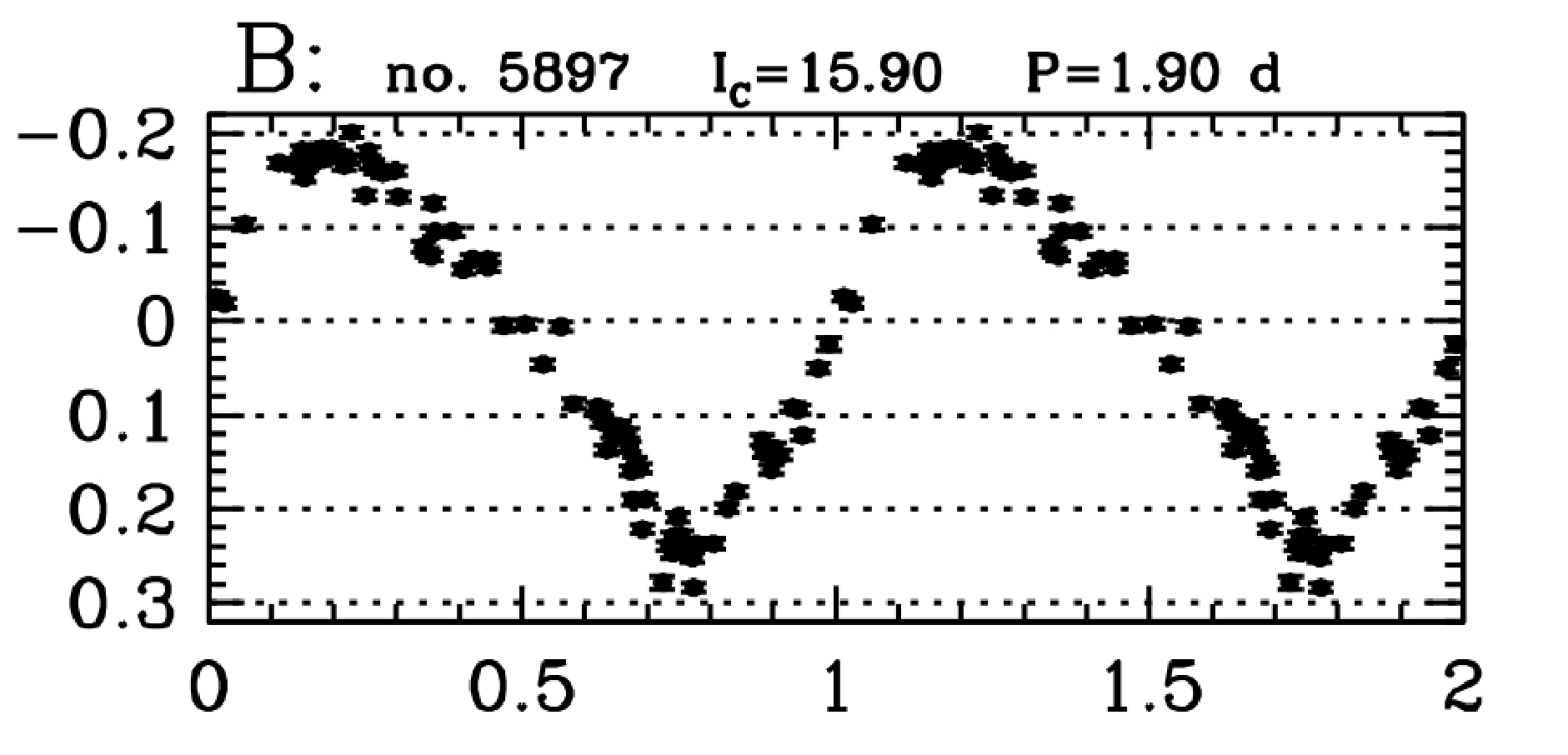}
   \vspace{0.5cm}
   \includegraphics[width=13cm]{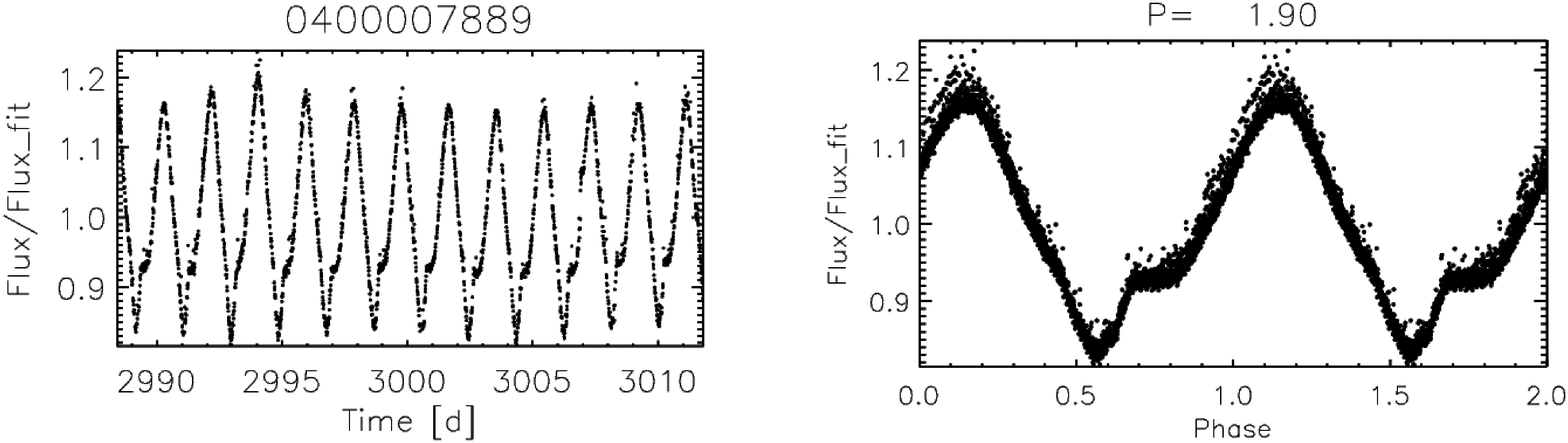}
      \caption{LC obtained from ground observations (\citealt{lbm+04}, upper left panel, between 30 Dec. 2000 and 01 Mar
      2001) and the folded light curve (upper right panel).
      CoRoT LC for the same star (lower left panel) and folded LC (lower right panel) The derived periods are indicated
      on the top of each folded LC panel, and are in agreement. Time in the CoRoT LC (bottom left panel)
      is in days from 2000 January 1 (JD-2451544.5).
              }
         \label{fig9}
   \end{figure*}

\begin{figure*}
   \centering
   \includegraphics[width=7cm]{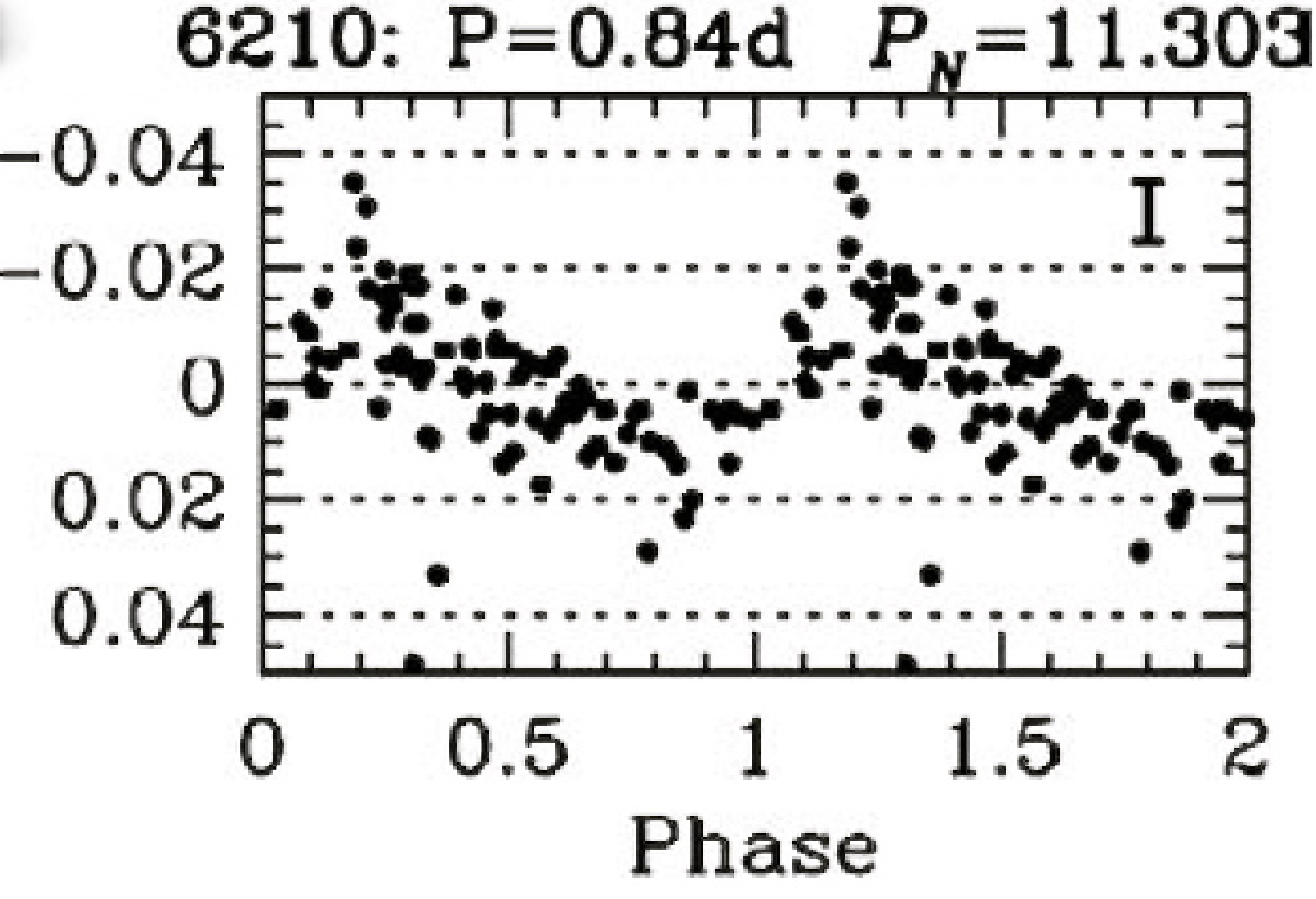}
    \vspace{0.5cm}
    \includegraphics[width=13cm]{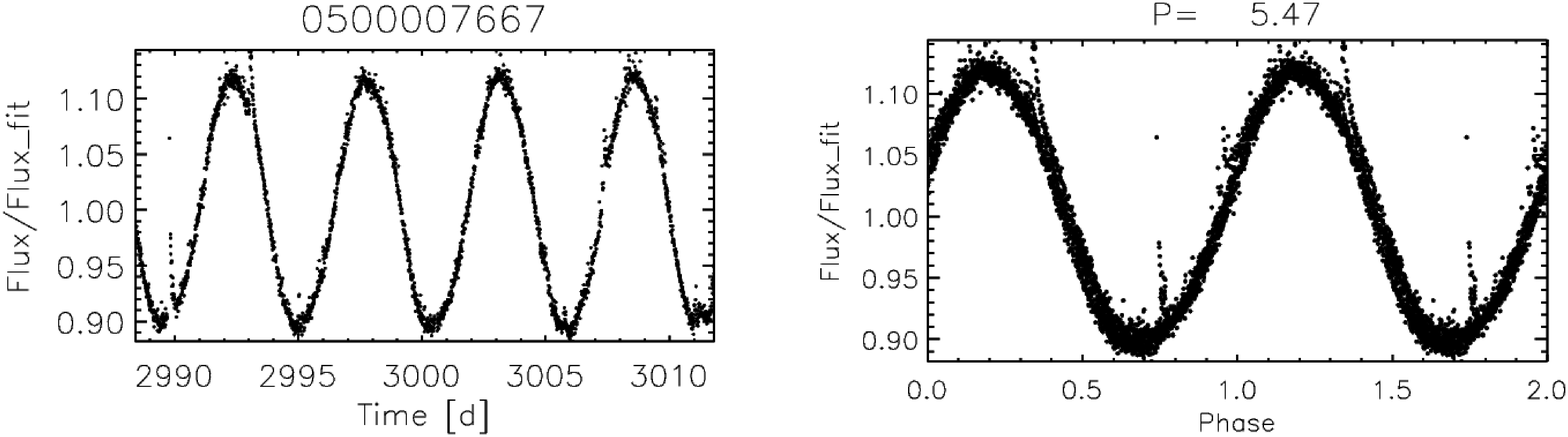}
      \caption{Folded LC from \citet{lbm+04} (top panel) with the CoRot LC and folded LC (bottom panel). In this case
      the ground determination of 0.84 d is clearly wrong, compared with the CoRoT determination of 5.5 d. Time in
      the CoRoT LC (bottom left panel) is in days from 2000 January 1 (JD-2451544.5).
              }
         \label{fig10}
   \end{figure*}

We now compare the periods derived from the CoRoT data with those of \citet{lbm+04,lmb+05}, who identified 405 periodic variables and 184 irregular variable members.
We analyzed the 103 NGC 2264 members in common with the \citet{lbm+04} ones. The CoRoT periods derived in the present study and ground derived ones (\citealt{lbm+04}), apart from the presence of few aliases, are
comparable with the exception of the 1-day periods derived from ground, likely affected by the day-night alternance (CoRoT LCs have both a better sampling and time
coverage than \citealt{lbm+04} ones).
A scatter plot of the CoRoT vs. ground periods is shown in Fig.~\ref{fig8}. We followed the indications of \citet{lbm+04} to estimate the
error in the measured periods, which is related to the finite frequency resolution $\delta \nu$ of the power spectrum. In our study we take advantage of the
uniformity of CoRoT sampling, thus the frequency resolution is related to the total time span T of the observations (23 d) with $\delta \nu \approx 1/T$\,
(\citealt{rld87}). Thus we assume as typical error in the period, that given by $\delta P \approx (\delta \nu P^{2})/2$. We found that for period shorter than 10 d,
the estimated error is $\le $\ 20\%, while it grows with period. With these indications, we find that 83\% (85/103) of the sample stars have consistent periods in the two
survey, most of the inconsistent values lie along either harmonics (9/103) or the 1-day aliases curve (5/103). \\We compare ground and CoRoT data for few intriguing cases, in Fig.~\ref{fig9} we reported a LC of a star with short period illustating the quality of the CoRoT LC with respect to the ground one. In Fig.~\ref{fig10} we show the folded LC from \citet{lbm+04} together with the CoRoT one, for a case in which the ground determination of 0.84 d is clearly wrong, in comparison with the CoRoT determination of 5.5 d. This comparison highlights the usefulness of a continuous temporal coverage, as can
only be achieved from space.\\

\section{Summary and conclusions}\label{concl}
We performed an accurate study of rotation periods in the young open cluster NGC 2264, based on the very accurate CoRoT photometry, also to search for a correlation between accretion and stellar rotation. 
We derived rotational period for 209 cluster members, out of 301, which are found to be periodic variables, with a
spot-like modulation of their light or AA Tau like.\\ This work enabled us to study the distribution of rotational periods of CTTSs and WTTSs. In order to
overcome the bias introduced by the use of NIR excess to classify CTTSs, we used $H\alpha$ equivalent widths. Furthermore the clean and continuous light
curves of CoRoT allowed us to avoid problems due to discontinuous time sampling. Thanks to the high quality of the CoRoT lightcurves, our results provide the most complete
set of rotational periods of NGC 2264 members, to date, compared to ground-based data.\\
As discussed in Sect. \ref{intro}, several models assume that accreting stars may be locked to their disks by strong magnetic fields that channel streams accretion (\citealt{sno+94}; \citealt{har02}). 
This would result in slower rotation velocities for CTTS, compared to WTTS which have no inner disks, and some observations suggested differences up to
a factor of two in rotational velocities of CTTS versus WTTS in Taurus (\citealt{esh+93}). However, several other studies have found no significant
variations between rotational velocities of CTTS and WTTS in ONC and IC 348 (\citealt{smm+99}, \citealt{rhm01}, \citealt{c&b06}). These results seem to
suffer from several biases affecting the selected samples and even the disk and accretion indicators adopted. These differences may also be due to a classification problem, on what one defines classical and weak line T-Tauri stars. 
\\
We found that the rotational distributions of CTTSs and WTTSs are different, with WTTSs rotating faster, with a median P$_{Rot}$ = 4.2 d, with respect to CTTSs,
with a median P$_{Rot}$ = 7.0 d. Our results are even more significant using a Class II - Class III comparison, probably because of the better
statistics. This could suggest that the presence of accretion, or any other properties related to star-disk interaction, affects the rotational period. It is
consistent with the disk locking scenario (\citealt{sno+94}; \citealt{har02}) and agrees with previous results in ONC (\citealt{c&h96};
\citealt{hbm+02}; \citealt{rsm+05}), also confirming the previous conclusions of \citet{lmb+05} and \citet{c&b07} for NGC 2264.\\
A new extensive multiband campaign has been executed on NGC 2264 in December 2011 (CoRoT+Spitzer+Chandra complemented with MOST observations). 
The new data will enable us to tackle some open questions, on the subject of this paper, such as the stability of the disk structure responsible for the AA Tau-like variability or differential rotation and the possibility of detecting period changes, when spots evolve and appear at different latitudes.

\begin{table*}
\begin{center}
\caption{Catalogue of periods for NGC 2264 members.}
\label{tab01}
\begin{tabular}{ccccrrrccc} \hline
      Corot ID	  & Period (d)$^a$ & AmpVar(\%)$^b$  &  R	 &   B-V  &RA(J2000.0)&DEC(J2000.0)   &Memb-IR$^c$ &EW(H$\alpha$)$^d$ &  Mass$^e$ \\\hline
     0223948127   &     I      &   20.7  &15.000 & -0.200 &  99.76562 &    9.67337    &   &	    &	       \\
     0223951807   &     -      &    0.9  &14.700 &  1.500 &  99.81503 &    9.65377    &   &	     &         \\
     0223951822   &     -      &    0.5  &13.800 &  0.690 &  99.81532 &    9.49160    &   &	     &         \\
     0223952236   &     -      &    4.6  &14.100 & -1.020 &  99.82101 &    9.97097    &   &	     &         \\
     0223953966   &   3.987    &    3.7  &12.600 & -0.590 &  99.84448 &    9.28458    &   &	     &         \\
     0223954040   &   9.684    &    2.7  &14.000 &  0.030 &  99.84557 &    9.60481    &   &	     &   1.30  \\
     0223954556   &     -      &    0.7  &13.800 &  0.400 &  99.85257 &    9.37606    &   &	     &         \\
     0223955032   &   5.436    &    5.3  &14.300 &  0.460 &  99.85937 &    9.68647    &   &   -1.100&	0.60   \\
     0223955438   &     -      &    0.7  &13.400 &  2.200 &  99.86499 &    9.38587    &   &	     &         \\
     0223955517   &     -      &    0.2  &12.900 &  0.520 &  99.86609 &    9.47768    &   &	     &         \\
     0223955994   &     -      &    0.7  &13.300 &  0.040 &  99.87251 &    9.34971    &   &	     &         \\
     0223956264   &   2.229    &    6.5  &14.800 &  0.470 &  99.87656 &    9.56053    &   &	     &   0.80	\\
     0223956963   &   9.786 AA &   15.6  &14.700 & -0.200 &  99.88675 &    9.07066    &   &	     &         \\
     0223957004   &     -      &    2.8  &14.200 &  0.700 &  99.88737 &    9.94127    &   &	     &         \\
     0223957142   &   2.568    &   16.9  &15.200 &  0.600 &  99.88912 &    9.86726    &   &	     &   0.40	\\
     0223957322   &   18.05    &    8.1  &15.300 &  0.850 &  99.89156 &    9.82254    &   &	     &   0.80	\\
     0223957455   &   10.16 AA &   28.0  &15.200 &  0.500 &  99.89333 &    9.91437    &   &	     &         \\
     0223957734   &     -      &    8.1  &15.300 &  0.690 &  99.89726 &    9.54231    &   &	     &         \\
     0223957908   &     -      &    0.7  &13.600 &  0.810 &  99.89991 &    9.40729    &   &	     &         \\
     0223958794   &     -      &    2.7  &14.200 &  0.730 &  99.91180 &    9.86451    &   &	     &   1.60  \\
     0223958963   &   0.859    &    9.0  &15.200 &  0.720 &  99.91379 &    9.93336    &   &	     &   0.60	\\
     0223959618   &   3.922 AA &   55.9  &14.200 &  0.860 &  99.92274 &    9.77229    &   &	     &         \\
     0223959652   &   3.732    &    5.7  &12.700 & -0.170 &  99.92316 &    9.57808    &   &   -3.300&	1.40  \\
     0223959949   &     -      &    0.6  &14.800 &  0.770 &  99.92764 &    9.53115    &   &	     &         \\
     0223960995   &     -      &    0.6  &14.500 &  1.310 &  99.94267 &    9.80571    &   &	     &         \\
     0223961132   &   3.839    &    9.3  &12.500 &  0.210 &  99.94466 &    9.68178    &III&	     &   1.40  \\
     0223961409   &   1.104    &    0.7  &11.900 & -0.030 &  99.94879 &    9.43535    &III&	     &         \\
     0223961560   &     -      &    1.4  &12.700 &  1.570 &  99.95090 &    9.98496    &   &	     &         \\
     0223961941   &   6.52     &    4.3  &14.700 &  0.400 &  99.95667 &    9.55630    &III&	     &   1.10  \\
     0223962024   &     -      &    0.7  &14.500 &  2.900 &  99.95769 &    9.93945    &   &	     &         \\
     0223962712   &     -      &    1.6  &14.500 &  0.970 &  99.96801 &    9.31938    &   &	     &         \\
     0223963678   &   0.676    &    5.0  &14.700 &  0.600 &  99.98199 &    9.79233    &III&	     &        \\
     0223963815   &     -      &    1.5  &13.200 &  0.100 &  99.98402 &    9.51299    &   &	     &        \\
     0223963881   &   12.92    &    0.7  &13.700 & -0.400 &  99.98499 &    9.72557    &III&	     &        \\
     0223963994   &     -      &    0.2  &15.200 & -1.200 &  99.98674 &    9.74040    &   &	     &        \\
     0223964667   &   6.456    &   79.3  &15.300 & -0.200 &  99.99690 &    9.45691    &   &	     &         \\
     0223964830   &   2.575    &    9.1  &15.900 &  0.530 &  99.99934 &    9.56164    &III&	     &   0.80	\\
     0223965280   &     -      &    0.7  &14.500 &  0.970 & 100.00603 &    9.51846    &   &	     &         \\
     0223965459   &   1.351    &   15.4  &15.600 &  0.700 & 100.00892 &    9.75401    &III&	     &         \\
     0223965593   &   9.830    &    0.8  &13.400 & -0.100 & 100.01107 &    9.30565    &III&	     &         \\
     0223965989   &   0.819    &    7.3  &15.300 &  0.760 & 100.01676 &    9.45211    &III&	     &   0.60	\\
     0223967301   &   0.957    &   12.6  &15.800 &  1.360 & 100.03550 &    9.73719    &III&    3.500&	0.50   \\
     0223967602   &   1.236    &    7.0  &13.300 &  0.470 & 100.04005 &    9.69555    &III&   -1.700&	1.50  \\
     0223967803   &   3.841    &    7.7  &14.300 &  0.600 & 100.04288 &    9.64872    &III&    2.000&	0.70   \\
     0223968039   &     I      &   56.9  &15.900 &  0.400 & 100.04640 &    9.63503    &   &    52.90&	0.90   \\
     0223968398   &   2.702    &   11.0  &15.800 &  1.130 & 100.05183 &    9.73986    &III&    3.600&	0.50   \\
     0223968439   &   8.688    &   31.6  &15.400 & -0.200 & 100.05240 &   10.09474    &II &	     &         \\
     0223968646   &     -      &    2.0  &15.400 &  0.500 & 100.05598 &    9.32465    &   &	     &         \\
     0223968688   &   1.117    &    3.9  &11.500 &  0.080 & 100.05666 &    9.41396    &III&	     &   2.20  \\
     0223968804   &   1.295    &    3.9  &13.600 &  0.660 & 100.05841 &    9.34141    &III&	     &   1.40  \\
     0223969098   &     I      &  158.2  &15.800 &  1.300 & 100.06316 &   10.03289    &   &	     &         \\
     0223969672   &     -      &    7.2  &16.000 &  1.000 & 100.07179 &   10.22651    &   &	     &         \\
     0223970440   &     -      &    3.7  &14.700 &  0.260 & 100.08372 &    9.47472    &   &    3.200&	   \\
     0223970694   &   1.467    &    5.2  &13.000 &  0.010 & 100.08755 &    9.60907    &III&    0.300&	1.50  \\
     0223971008   &   7.38     &    9.9  &15.900 &  1.800 & 100.09258 &    9.90811    &III&    2.100&	0.60   \\
     0223971231   &     I      &  170.8  &14.500 & -0.340 & 100.09618 &    9.46190    &   &    49.50&	1.10  \\
     0223971383   &   4.648    &   11.9  &15.800 &  0.550 & 100.09904 &    9.92345    &II &    80.60&	   \\
     0223971866   &   7.015    &    2.7  &12.600 &  0.360 & 100.10703 &    9.97667    &III&	     &         \\
     0223971984   &   6.281    &   14.0  &15.500 &  1.480 & 100.10938 &    9.63385    &III&    3.500&	0.60   \\
     0223972652   &     I      &   21.4  &13.000 & -0.900 & 100.11988 &    9.51704    &   &    39.10&	   \\
     0223972652   &     -      &   21.4  &13.000 & -0.900 & 100.11988 &    9.51704    &   &    1.800&	   1.60  \\\hline
\end{tabular}
\end{center}
\end{table*}

\addtocounter{table}{-1}
\begin{table*}
\caption{Continued.}
\begin{center}
\begin{tabular}{ccccrrrccc} \hline
      Corot ID	  & Period (d)$^a$ &   AmpVar(\%)$^b$ & R &   B-V  &    RA(J2000.0)&  DEC(J2000.0) &Memb-IR$^c$  &EW(H$\alpha$)$^d$ &  Mass$^e$ \\\hline
     0223972691   &   7.206    &    12.0  &15.300 &  1.310 & 100.12061 &    9.70490    &III&	1.700&   0.90	\\
     0223972918   &     -      &     0.9  &14.100 &  0.930 & 100.12383 &    9.99277    &   &	       &	 \\
     0223972960   &     -      &     2.5  &15.300 &  2.370 & 100.12453 &    9.83637    &   &	       &   1.60  \\
     0223973200   &     I      &   113.2  &15.600 &  0.070 & 100.12858 &    9.57811    &   &	22.20&   1.30  \\
     0223973292   &   1.974    &    17.7  &14.700 &  0.170 & 100.13010 &    9.51878    &III&	1.700&   0.90	\\
     0223973318   &     I      &     1.0  &15.600 &  2.600 & 100.13048 &    9.69152    &   &	       &	 \\
     0223973692   &   3.456    &    14.6  &15.700 &  1.760 & 100.13673 &    9.85833    &   &	2.700	&   0.90   \\
     0223974593   &   1.156    &    18.1  &14.600 &  0.580 & 100.15134 &    9.31610    &III&	       &   0.50   \\
     0223974689   &     -      &     1.8  &14.800 &  1.100 & 100.15278 &    9.36818    &   &	       &	 \\
     0223974891   &   1.212    &     2.9  &12.900 &  1.740 & 100.15620 &    9.91617    &III&	       &   2.10  \\
     0223975203   &   9.786    &     1.6  &15.200 &  0.200 & 100.16160 &    9.36077    &III&	       &   0.90   \\
     0223975253   &     -      &    10.5  &12.600 &  1.440 & 100.16248 &    9.60013    &   &	3.500&   0.60	\\
     0223975844   &   3.332    &     8.9  &15.200 &  3.130 & 100.17236 &    9.90389    &II &	12.20&   1.50  \\
     0223976028   &     I      &     4.1  &13.400 &  0.710 & 100.17587 &    9.56050    &   &	7.300&   1.70  \\
     0223976099   &   14.17    &     5.6  &15.800 &  0.540 & 100.17683 &    9.53921    &III&	0.500&   1.20  \\
     0223976494   &   2.267    &    12.7  &14.300 &  0.800 & 100.18377 &    9.39887    &III&	       &   1.10  \\
     0223976672   &   15.00    &     5.2  &14.100 &  1.540 & 100.18687 &    9.96244    &III&	1.600&   0.90	\\
     0223976747   &   3.173 AA &    96.6  &14.700 &  0.750 & 100.18816 &    9.47915    &   &	7.200&   1.50  \\
     0223977051   &   4.53     &     5.5  &15.200 &  1.410 & 100.19336 &    9.99640    &   &	       &	 \\
     0223977092   &     -      &     0.5  &14.000 &  2.000 & 100.19403 &    9.36149    &   &	       &	 \\
     0223977232   &   0.712    &     0.8  &14.300 &  0.340 & 100.19633 &    9.30929    &III&	       &	 \\
     0223977953   &   4.919 AA &    26.6  &15.700 &  0.730 & 100.20782 &    9.61389    &II &	66.30&   1.40  \\
     0223978227   &   3.779    &    20.7  &16.000 &  2.890 & 100.21194 &    9.93148    &III&	2.300&   0.80	\\
     0223978308   &   5.374 AA &     5.6  &15.400 &  3.650 & 100.21328 &    9.74633    &II &	3.500&   1.80  \\
     0223978921   &     I      &     5.4  &15.800 &  1.400 & 100.22346 &    9.55701    &   &	18.20&   1.40  \\
     0223978947   &   8.5      &     0.2  &13.600 &  1.340 & 100.22403 &    9.51095    &III&	       &   1.80  \\
     0223979728   &     I      &     8.3  &15.800 &  1.410 & 100.23665 &    9.63043    &   &	113.2&   1.10  \\
     0223979759   &   3.84     &     0.6  &15.400 &  3.770 & 100.23719 &    9.81144    &III&	3.900&   1.60  \\
     0223979980   &   0.577    &     6.9  &14.800 &  2.200 & 100.24097 &    9.94176    &III&	       &	 \\
     0223980019   &     I      &     0.7  &14.500 & -0.500 & 100.24154 &    9.30101    &   &	       &	 \\
     0223980048   &   12.5  AA &    24.4  &14.100 &  1.840 & 100.24200 &    9.61504    &II &	34.00&      \\
     0223980233   &     -      &    14.0  &13.800 &  1.800 & 100.24457 &    9.60384    &   &	22.20&   0.30	\\
     0223980258   &   6.990    &     8.9  &14.900 &  0.980 & 100.24509 &    9.65531    &II &	27.90&   0.60	\\
     0223980264   &   3.482 AA &    54.0  &14.600 &  0.960 & 100.24516 &    9.51607    &II &	14.30&   1.90  \\
     0223980412   &   6.39     &    12.1  &14.800 &  0.760 & 100.24764 &    9.99601    &II &	7.400&   1.10  \\
     0223980447   &   1.675    &     9.2  &16.000 &  1.220 & 100.24816 &    9.58649    &II &	6.400&   0.90	\\
     0223980621   &   3.049    &     8.9  &13.400 &  1.770 & 100.25099 &    9.98056    &III&   -1.200&   1.80  \\
     0223980688   &     I      &    72.6  &16.000 &  1.310 & 100.25205 &    9.75101    &   &	15.00&   1.30  \\
     0223980693   &   5.282 AA &    112.0 &14.900 &  1.240 & 100.25214 &    9.48791    &II &	16.60&   1.50  \\
     0223980807   &     I      &     8.1  &14.600 &  1.790 & 100.25407 &    9.54585    &   &	6.400&      \\
     0223980941   &   3.794    &    85.9  &15.300 &  1.800 & 100.25637 &   10.24905    &   &	       &	 \\
     0223980988   &   8.58     &     1.6  &15.400 &  0.000 & 100.25703 &    9.35170    &III&	       &	 \\
     0223981023   &   7.320 AA &    38.2  &15.800 &  1.300 & 100.25770 &    9.64490    &   &	1.500&   1.50  \\
     0223981174   &   1.974    &    19.7  &15.600 &  1.000 & 100.26061 &    9.58235    &III&	0.600&   1.50  \\
     0223981250   &   7.437    &    32.8  &14.800 & -0.200 & 100.26187 &   10.12015    &   &	       &	 \\
     0223981285   &   1.152    &     1.0  &15.300 &  2.300 & 100.26239 &    9.79856    &III&	       &	 \\
     0223981349   &   8.014    &    13.4  &15.000 &  2.300 & 100.26363 &    9.96535    &III&	1.500&   0.80	\\
     0223981406   &   2.157    &     9.0  &13.800 &  1.190 & 100.26449 &    9.52188    &III&   -2.500&   1.60  \\
     0223981535   &   4.557    &    15.1  &15.400 &  2.100 & 100.26640 &    9.96940    &III&	4.100&      \\
     0223981550   &   14.58    &     0.5  &14.900 & -2.200 & 100.26661 &    9.39267    &   &	       &   1.20  \\
     0223981753   &   2.971    &     1.1  &13.000 &  1.070 & 100.26965 &    9.60751    &III&	       &   2.50  \\
     0223981811   &     I      &    86.4  &15.900 &  4.140 & 100.27067 &    9.84631    &   &	36.50&   1.60  \\
     0223982076   &   2.468    &     3.3  &13.000 &  2.440 & 100.27471 &    9.45502    &III&	2.600&   0.30	\\
     0223982136   &   3.018    &    10.7  &14.700 &  1.280 & 100.27571 &    9.60653    &   &	10.00&   1.40  \\
     0223982169   &   3.162    &    10.3  &14.900 &  1.590 & 100.27621 &    9.49197    &III&	3.100&   0.30	\\
     0223982299   &   4.671 AA &    15.1  &14.700 &  0.400 & 100.27850 &    9.03797    &   &	       &	 \\
     0223982375   &   3.320    &    13.6  &15.400 & -0.200 & 100.27966 &    9.21076    &II &	       &	 \\
     0223982407   &   2.582    &    23.6  &14.600 &  1.310 & 100.28017 &    9.97540    &III&	1.400&   1.10  \\
     0223982423   &   9.026    &     3.8  &15.800 &  2.400 & 100.28040 &   10.22570    &   &	       &	 \\
     0223982535   &   5.052    &     4.7  &14.700 &  1.310 & 100.28240 &    9.73427    &III&	       &	 \\
     0223982779   &   1.882    &    26.8  &15.600 & -2.000 & 100.28673 &    9.39554    &III&	2.400&   0.60	\\\hline
\end{tabular}
\end{center}
\end{table*}

\addtocounter{table}{-1}
\begin{table*}
\caption{Continued.}
\begin{center}
\begin{tabular}{ccccrrrccc} \hline
      Corot ID	  & Period (d)$^a$ &   AmpVar(\%)$^b$  & R &   B-V  &    RA(J2000.0)&DEC(J2000.0)  &Memb-IR$^c$  &EW(H$\alpha$)$^d$ &  Mass$^e$ \\\hline
     0223982807   &     -      &     2.3   &13.400 &  5.000 & 100.28716 &   10.23981   & III&	      & 	\\
     0223983310   &   3.589    &     0.9   &13.800 &  0.510 & 100.29544 &   10.01147   & III&	      & 	\\
     0223983509   &   2.390    &     7.89  &14.100 &  0.300 & 100.29829 &   10.04005   &    &	      & 	\\
     0223983925   &   3.704    &    11.29  &15.500 &  2.000 & 100.30511 &    9.91922   &    &	3.000&      \\
     0223984075   &   3.793    &     3.6   &12.700 &  1.800 & 100.30750 &    9.92897   & III&	-2.600&      \\
     0223984253   &   10.42    &     9.6   &16.000 &  1.530 & 100.31026 &    9.55614   & III&	1.900&   1.10  \\
     0223984520   &   1.469    &     7.2   &15.200 &  1.590 & 100.31425 &    9.77779   & III&	0.200&   1.70  \\
     0223984572   &     I      &    14.5   &12.300 &  2.400 & 100.31499 &    9.44282   & III&	5.500&   0.30	\\
     0223984572   &     -      &    14.5   &12.300 &  2.400 & 100.31499 &    9.44282   & III&		 &	   \\
     0223984600   &   5.343    &     1.0   &13.600 & -0.300 & 100.31541 &    9.63857   &    &		&	  \\
     0223984608   &   6.098    &     7.7   &10.200 &  3.180 & 100.31551 &    9.43795   &    &	 1.900&    1.20  \\
     0223985009   &     I      &    83.4   &15.600 &  2.360 & 100.32182 &    9.90918   & III&	 58.30&    0.80   \\
     0223985176   &   6.547    &     8.3   &15.700 &  1.240 & 100.32470 &    9.56046   & III&	 2.900&    0.60   \\
     0223985261   &   18.08 AA &    20.3   &15.400 &  1.280 & 100.32613 &    9.56501   &    &	 28.90&    1.40  \\
     0223985611   &   4.94     &    21.5   &14.900 &  1.270 & 100.33174 &    9.52915   & III&	 1.500&    1.10  \\
     0223985845   &   2.604    &     9.4   &15.900 &  1.220 & 100.33559 &    9.75999   &    &		&	  \\
     0223985987   &   3.308 AA &    44.3   &15.300 &  1.300 & 100.33751 &    9.56029   & III&	 10.60&    0.90   \\
     0223986498   &   3.206    &     6.5   &14.800 &  1.680 & 100.34600 &    9.45753   & III&		&	  \\
     0223986686   &     -      &     0.6   &11.900 &  0.090 & 100.34904 &    9.56587   & II &		&	  \\
     0223986811   &   7.92     &     2.7   &15.700 &  1.020 & 100.35109 &    9.53181   & III&		&	  \\
     0223986923   &   8.300    &     0.8   &15.200 &  1.160 & 100.35297 &    9.43999   &    &		&	  \\
     0223987178   &   9.84 AA  &    36.8   &15.600 &  0.850 & 100.35670 &    9.57878   & III&	 15.90&    0.60   \\
     0223987553   &   1.544    &    22.3   &14.300 &  0.480 & 100.36308 &    9.58516   & III&	 1.100&    1.50  \\
     0223987997   &   6.456    &    22.9   &13.800 &  0.020 & 100.36986 &    9.64432   & II &		&	  \\
     0223988020   &     -      &     1.0   &14.200 & -0.200 & 100.37021 &   10.15428   & III&		&	  \\
     0223988099   &     -      &     2.5   &13.700 &  0.660 & 100.37155 &    9.66014   & II &	-1.400&    1.60  \\
     0223988099   &   3.273    &     2.5   &13.700 &  0.660 & 100.37155 &    9.66014   &    &	-1.400&    1.60  \\
     0223988742   &   5.025    &    16.7   &15.300 &  1.270 & 100.38172 &    9.80926   &    &	 5.200&    1.60  \\
     0223988827   &   4.767    &     9.3   &15.800 &  0.810 & 100.38329 &   10.00690   & III&	 13.10&    1.10  \\
     0223988965   &   9.5      &     4.5   &14.200 &  0.640 & 100.38538 &    9.63550   & III&	 1.300&    0.90   \\
     0223989567   &     I      &     8.4   &15.200 &  1.700 & 100.39403 &    9.60913   & III&	 4.500&    0.50   \\
     0223989989   &   6.547    &     7.2   &15.900 &  0.800 & 100.40102 &    9.65579   & III&	 1.600&    0.80   \\
     0223990299   &   4.469    &     9.3   &14.500 &  0.660 & 100.40536 &    9.75196   &    &	 35.00&    1.20  \\
     0223990764   &     -      &     0.6   &15.800 &  3.200 & 100.41270 &    9.49399   & III&		&	  \\
     0223990964   &   10.17 AA &    11.5   &13.600 &  0.340 & 100.41553 &    9.67456   & II &	 52.50&    1.60  \\
     0223991355   &     -      &     0.9   &12.700 &  0.000 & 100.42163 &    9.54543   &    &		&	  \\
     0223991789   &   3.956    &    11.7   &14.600 &  0.560 & 100.42803 &    9.71584   & II &	 0.900&    0.60   \\
     0223991832   &   8.608    &    18.2   &16.000 & -0.100 & 100.42868 &    9.41913   &    &		&	  \\
     0223991967   &     -      &     0.1   &12.500 &  0.650 & 100.43058 &    9.45033   & III&		&	  \\
     0223992383   &   3.380    &     5.1   &15.200 &  0.310 & 100.43725 &    9.74473   & II &		&   0.80   \\
     0223992685   &   18.5     &     0.7   &12.300 &  0.300 & 100.44153 &    9.39736   &    &		&	  \\
     0223993084   &   6.456    &     5.3   &15.400 & -0.140 & 100.44751 &    9.70001   & III&	 2.600&    0.50   \\
     0223993180   &   2.411    &     1.4   &12.900 &  0.430 & 100.44890 &    9.86750   & III&		&   1.80  \\
     0223993199   &     I      &    40.3   &14.200 &  0.740 & 100.44914 &    9.56957   & III&		&	  \\
     0223993277   &   1.184    &     3.9   &13.200 &  0.110 & 100.45027 &    9.71222   & III&		&   1.50  \\
     0223993499   &     I      &     0.9   &14.500 &  3.200 & 100.45351 &    9.72045   &    &		&   1.70  \\
     0223993840   &   3.250    &    35.8   &14.500 &  1.050 & 100.45837 &    9.49241   & III&		&   0.90   \\
     0223994268   &   3.762    &     4.9   &13.200 &  0.240 & 100.46436 &    9.89531   &    &		&   1.50  \\
     0223994721   &     I      &    37.9   &14.200 &  0.110 & 100.47102 &    9.96762   & III&		&   0.80   \\
     0223994760   &   5.634    &     8.9   &14.800 &  0.580 & 100.47147 &    9.84660   & III&	 1.800&    1.10  \\
     0223995167   &     -      &     0.4   &13.300 &  2.000 & 100.47691 &    9.48783   &    &		&	  \\
     0223995308   &   10.5 AA  &    53.6   &15.800 & -0.130 & 100.47881 &    9.71481   & II &		&   0.30   \\
     0223995327   &     -      &     0.8   &13.200 &  0.230 & 100.47910 &    9.89077   &    &		&	  \\
     0223997570   &   3.660    &     6.5   &14.500 &  0.670 & 100.51064 &    9.61480   &    &		&	  \\
     0223997608   &     -      &     2.7   &14.700 &  0.260 & 100.51109 &    9.97452   &    &		&	  \\
     0223998980   &     -      &     2.2   &15.500 &  0.600 & 100.52973 &    9.89573   & III&		&	  \\
     0223999063   &     -      &     0.3   &11.700 &  0.410 & 100.53070 &    9.82991   &    &		&	  \\
     0223999581   &     -      &     0.5   &14.200 &  1.220 & 100.53790 &    9.98422   &    &		&	  \\
     0223999591   &     -      &     1.2   &15.700 &  0.500 & 100.53800 &    9.80151   &    &		&	  \\
     0224000646   &     -      &     0.5   &13.700 &  0.900 & 100.55241 &    9.98546   &    &		&	  \\
     0224000835   &     -      &     1.3   &15.700 & -0.400 & 100.55477 &    9.76773   &    &		&	  \\
     0224001158   &     I      &     5.8   &15.700 & -0.300 & 100.55867 &    9.59576   &    &		&	  \\\hline

\end{tabular}
\end{center}
\end{table*}

\addtocounter{table}{-1}
\begin{table*}
\caption{Continued.}
\begin{center}
\begin{tabular}{ccccrrrccc} \hline
      Corot ID	  & Period (d)$^a$ & AmpVar(\%)$^b$   &R  &   B-V  &   RA(J2000.0) &DEC(J2000.0)   &Memb-IR$^c$  &EW(H$\alpha$)$^d$ &  Mass$^e$ \\\hline
     0224001312   &     -      &     1.0  &14.200 & -0.150 & 100.56084 &    9.84115    &    &	       &	 \\
     0224003566   &     -      &     0.5  &14.700 & -0.200 & 100.59132 &    9.80927    &    &	       &	 \\
     0224006123   &   10.25 AA &    53.2  &13.000 &  0.400 & 100.62704 &    9.15735    &    &	       &	 \\
     0400007328   &   2.434    &    23.0  &13.530 &  2.450 & 100.32380 &    9.49061    & III&	 2.400&   0.30   \\
     0400007394   &   3.443    &     4.6  &14.020 &  3.020 & 100.21672 &    9.75134    &    &	 2.800&   0.50   \\
     0400007528   &   9.42     &    14.0  &14.450 &  1.860 & 100.15780 &    9.58167    & II &	 23.40&   0.30   \\
     0400007529   &   4.842    &     9.3  &14.560 &  2.250 & 100.21948 &    9.73917    & III&	 2.200&   0.40   \\
     0400007538   &     I      &    23.1  &14.450 &  0.230 & 100.15217 &    9.84601    &    &	 21.10&   0.40   \\
     0400007614   &     I      &    13.6  &14.640 &  1.580 & 100.05709 &    9.94183    &    &	 130.2&   0.40   \\
     0400007686   &     I      &     5.1  &15.150 & -8.100 & 100.27679 &    9.47745    &    &	 56.10&   0.40   \\
     0400007687   &   11.5     &     5.6  &15.220 &  1.700 & 100.30544 &    9.86512    & III&	 2.000&   0.50   \\
     0400007702   &   5.884    &    20.2  &15.230 &  1.550 & 100.15916 &    9.49792    & II &	 2.600&   0.50   \\
     0400007709   &     -      &    10.3  &14.950 &  1.720 & 100.35226 &    9.62654    &    &	 8.900&   0.30   \\
     0400007734   &   9.996    &     5.6  &15.200 &  1.830 & 100.36250 &    9.50365    &    &	 25.80&   0.50   \\
     0400007743   &     -      &     0.9  &15.230 &  1.910 & 100.37020 &    9.58169    &    &	       &	 \\
     0400007765   &     I      &     2.7  &15.300 &  1.840 & 100.23683 &    9.86573    &    &	 1.100&   0.30   \\
     0400007784   &   9.114    &     6.1  &15.350 &  1.480 & 100.00467 &    9.59265    & III&	       &	 \\
     0400007786   &   8.608    &     6.7  &15.460 &  1.180 & 100.21748 &    9.94537    &    &	 3.100&   0.50   \\
     0400007803   &   9.73     &    12.3  &14.680 &  1.130 & 100.26503 &    9.50806    & II &	 20.40&      \\
     0400007809   &   3.990    &    14.7  &15.510 &  1.510 & 100.12186 &    9.73542    &    &	 31.30&   0.40   \\
     0400007860   &   2.172    &     1.8  &15.280 &  0.810 & 100.23959 &    9.82246    & III&	       &	 \\
     0400007889   &   1.897    &    46.3  &15.190 &  1.810 & 100.27310 &    9.52793    &    &	 3.000&   0.30   \\
     0400007919   &   4.625    &    15.0  &15.620 &  1.600 & 100.26164 &    9.38756    &    &	       &	 \\
     0400007955   &     I      &    17.1  &15.900 &  1.380 & 100.21982 &    9.71679    &    &	 19.40&   0.30   \\
     0400007956   &   1.260    &     3.6  &15.330 &  1.580 & 100.27903 &    9.68180    & III&	 2.800&      \\
     0400007957   &     I      &    16.9  &15.820 &  1.680 & 100.32107 &    9.54786    &    &	 2.100&   0.30   \\
     0400007959   &   5.738    &    15.5  &15.600 &  0.970 & 100.27805 &    9.79100    & II &	 6.200&      \\
     0400008031   &     -      &    29.3  &15.840 &  1.110 & 100.26287 &    9.48460    &    &	 14.60&      \\
     0400008086   &   5.34     &     3.4  &15.890 &  1.000 & 100.23363 &    9.71502    &    &	 8.000&      \\
     0400008126   &   0.546    &     1.7  &15.540 &  0.970 & 100.29521 &    9.88840    & III&	 13.20&      \\
     0500007008   &     -      &     5.3  &10.350 &  1.180 & 100.15522 &    9.79159    &    &	       &	 \\
     0500007018   &   1.487    &     1.0  &10.950 &  1.360 & 100.02357 &    9.59702    & III&	       &	 \\
     0500007021   &     -      &     0.3  &10.720 &  1.010 & 100.48482 &    9.83499    &    &	       &	 \\
     0500007022   &   3.332    &     5.6  &11.050 &  1.310 & 100.30433 &    9.45886    & III&	       &	 \\
     0500007025   &   0.747    &     7.9  &11.360 &  1.230 & 100.19200 &    9.82149    & III&	       &   3.00  \\
     0500007031   &     -      &     0.8  &11.450 &  0.820 & 100.19658 &    9.48052    &    &	       &	 \\
     0500007038   &   4.132    &     0.8  &11.880 &  1.060 & 100.15281 &    9.78959    & III&	       &	 \\
     0500007039   &   11.92    &     0.6  &11.980 &  1.140 & 100.27870 &    9.38927    & III&	-1.500&   2.20  \\
     0500007046   &     I      &     3.6  &12.550 &  1.470 & 100.18600 &    9.80059    &    &	 48.90&      \\
     0500007051   &   10.0     &     1.1  &11.890 &  0.700 & 100.25919 &    9.86443    & III&	-1.600&   2.10  \\
     0500007087   &   14.0     &     2.0  &12.340 &  0.360 & 100.09639 &    9.93886    & III&	       &	 \\
     0500007089   &     I      &    21.4  &12.320 &  0.260 & 100.30362 &    9.43746    &    &	 85.60&   1.70  \\
     0500007115   &   1.995    &    35.5  &13.420 &  1.050 & 100.30241 &    9.87533    & II &	 35.30&      \\
     0500007120   &   8.53     &    14.2  &13.450 &  1.010 & 100.19793 &    9.82471    & II &	 12.80&   1.80  \\
     0500007122   &     I      &    57.2  &12.780 &  0.290 & 100.37966 &    9.44951    &    &	 25.90&      \\
     0500007126   &     -      &     0.6  &12.550 & -0.040 & 100.23119 &    9.52272    &    &	       &	 \\
     0500007137   &   2.914    &     7.6  &12.780 &  0.020 & 100.29095 &    9.45339    & III&	 3.100&   1.20  \\
     0500007157   &   4.344    &     7.2  &13.190 &  0.260 & 100.25324 &    9.85620    & III&	 1.600&   1.30  \\
     0500007176   &   4.024    &     6.9  &13.400 &  0.330 & 100.26849 &    9.85725    & III&	       &	 \\
     0500007197   &   9.114    &     5.0  &13.570 &  0.220 & 100.18063 &    9.84988    & III&	 1.700&   0.80   \\
     0500007202   &     -      &     0.3  &13.650 &  0.260 & 100.20934 &    9.33399    &    &	       &	 \\
     0500007209   &     I      &    45.2  &13.600 &  0.170 & 100.21081 &    9.91593    &    &	 11.20&   1.60  \\
     0500007217   &   2.582    &    10.4  &13.520 &  0.000 & 100.27172 &    9.88772    & III&	 1.500&      \\
     0500007221   &     I      &     9.3  &13.430 & -0.130 & 100.29939 &    9.44207    &    &	 5.400&   0.70   \\
     0500007227   &   7.151    &     6.9  &13.880 &  0.280 & 100.12758 &    9.76962    & III&	 1.500&   1.30  \\
     0500007248   &   7.50     &     2.5  &13.860 &  0.170 & 100.27125 &    9.86238    & III&	 1.700&   1.10  \\
     0500007249   &     I      &     5.9  &14.680 &  0.980 & 100.41155 &    9.53661    &    &	 58.60&   1.20  \\
     0500007252   &   13.88    &    39.8  &13.730 &  0.000 & 100.16299 &    9.84962    & II &	 46.50&   1.50  \\
     0500007269   &   3.674    &    92.7  &14.530 &  0.740 & 100.17435 &    9.86237    & II &	 22.90&   1.10  \\
     0500007272   &   3.748    &    14.9  &13.380 & -0.420 & 100.16840 &    9.84735    & III&	 58.30&   0.70   \\
     0500007276   &   4.743    &     7.2  &13.970 &  0.150 & 100.17261 &    9.80267    &    &	 2.400&   1.10  \\
     0500007283   &   3.217    &    9.6   &13.940 &  0.100 & 100.23215 &    9.85385    &II &	8.000&   1.00  \\\hline
\end{tabular}
\end{center}
\end{table*}

\addtocounter{table}{-1}
\begin{table*}
\caption{Continued.}
\begin{center}
\begin{tabular}{ccccrrrccc} \hline
      Corot ID	  & Period (d)$^a$ &AmpVar(\%)$^b$    & R &   B-V  &  RA(J2000.0)  &DEC(J2000.0)   &Memb-IR$^c$  &EW(H$\alpha$)$^d$ &  Mass$^e$ \\\hline
     0500007298   &   15.25    &    4.3   &14.070 &  0.150 & 100.15262 &    9.80638    &III&	4.900&   1.10  \\
     0500007300   &     I      &    0.7   &13.480 & -0.450 & 100.15151 &    9.37904    &   &	      & 	\\
     0500007308   &   3.141    &    9.3   &14.200 &  0.230 & 100.10616 &    9.80721    &III&	0.800&   1.10  \\
     0500007315   &   7.812 AA &   32.9   &13.920 & -0.080 & 100.17216 &    9.85066    &II &	24.50&   0.80	\\
     0500007330   &   4.304    &    3.5   &14.030 & -0.030 & 100.27422 &    9.87996    &III&	0.900&   0.80	\\
     0500007335   &   14.99 AA &    8.8   &14.020 & -0.070 & 100.26789 &    9.41449    &II &	101.8&   0.60	\\
     0500007347   &   4.206    &   15.3   &14.210 &  0.060 & 100.25000 &    9.48057    &III&	1.700&   0.80	\\
     0500007354   &   1.165    &   11.6   &14.240 &  0.050 & 100.22991 &    9.84718    &   &	2.800&   1.00  \\
     0500007366   &   3.617    &    8.2   &14.060 & -0.190 & 100.19733 &    9.81373    &III&	1.800&   0.50	\\
     0500007369   &     I      &   15.2   &13.590 & -0.680 & 100.27808 &    9.57943    &   &	49.40&      \\
     0500007379   &   14.17    &    2.7   &14.300 & -0.010 & 100.17095 &    9.79936    &   &	7.500&   0.80	\\
     0500007383   &   1.289    &    4.8   &13.970 & -0.360 & 100.27368 &    9.90520    &III&	3.800&   0.50	\\
     0500007410   &     -      &    2.8   &14.050 & -0.390 & 100.21897 &    9.86833    &   &	6.900&   0.70	\\
     0500007416   &   3.748    &    3.5   &14.170 & -0.300 & 100.27124 &    9.81332    &III&	      & 	\\
     0500007457   &   1.049    &    3.0   &14.480 & -0.130 & 100.17415 &    9.83120    &III&	4.200&   0.60	\\
     0500007458   &   4.625    &    5.8   &14.490 & -0.130 & 100.18768 &    9.76162    &   &	2.400&   0.80	\\
     0500007460   &   8.49 AA  &   42.0   &14.600 & -0.030 & 100.18006 &    9.78535    &II &	27.10&   0.90	\\
     0500007473   &     I      &   24.9   &14.430 & -0.260 & 100.22610 &    9.82232    &   &	161.1&   0.50	\\
     0500007505   &     I      &    8.3   &14.900 &  0.070 & 100.17086 &    9.46509    &   &	13.20&      \\
     0500007550   &     -      &    0.4   &14.590 & -0.440 & 100.40549 &    9.53271    &   &	      & 	\\
     0500007556   &   3.405    &    4.5   &14.250 & -0.810 & 100.34229 &    9.35863    &III&	3.800&      \\
     0500007572   &   8.08     &    3.2   &14.440 & -0.670 & 100.29298 &    9.36376    &III&	2.700&      \\
     0500007580   &   1.805    &   17.5   &14.700 & -0.450 & 100.24931 &    9.86359    &III&	4.900&   0.40	\\
     0500007585   &   9.786    &   14.3   &14.780 & -0.410 & 100.19170 &    9.29951    &   &	      & 	\\
     0500007610   &   9.34     &   17.4   &14.690 & -0.610 & 100.24792 &    9.49770    &II &	26.20&   0.30	\\
     0500007634   &   11.25    &    6.7   &14.970 & -0.410 & 100.26488 &   10.00983    &   &	6.900&   0.50	\\
     0500007667   &   5.405    &   22.6   &15.190 & -0.340 & 100.31035 &    9.62065    &III&	4.100&      \\
     0500007682   &     I      &    7.9   &14.580 & -1.010 & 100.31008 &    9.44952    &   &	2.700&   0.30	\\
     0500007708   &   9.584    &   11.4   &15.220 & -0.450 & 100.22546 &    9.49752    &III&	3.600&      \\
     0500007727   &     I      &   20.7   &15.150 & -0.590 & 100.29583 &    9.59881    &   &	61.50&   0.80	\\
     0500007730   &   12.5     &    3.2   &14.850 & -0.910 & 100.20505 &    9.96077    &   &	50.80&   0.50	\\
     0500007752   &   4.042    &    5.4   &14.720 & -1.110 & 100.28734 &    9.56278    &II &	51.00&   0.30	\\
     0500007766   &     I      &    7.2   &15.070 & -0.800 & 100.29283 &    9.55696    &   &	2.700&   0.30	\\
     0500007770   &   10.0     &    4.4   &15.300 & -0.580 & 100.01115 &    9.69690    &III&	      & 	\\
     0500007794   &   8.854    &    7.1   &15.410 & -0.530 & 100.23939 &    9.48984    &III&	4.000&      \\
     0500007808   &   5.025    &    2.9   &15.060 & -0.920 & 100.29496 &    9.77811    &III&	3.900&   0.30	\\
     0500007816   &   7.378    &    1.1   &15.060 & -0.980 & 100.22344 &    9.78455    &II &	8.100&   0.30	\\
     0500007837   &     I      &   19.5   &15.380 & -0.740 & 100.28690 &    9.88365    &   &	      & 	\\
     0500007857   &     I      &    8.0   &15.360 & -0.820 & 100.26905 &    9.64190    &   &	108.0&   0.30	\\
     0500007874   &     I      &    3.3   &15.740 & -0.500 & 100.18720 &    9.81921    &   &	2.600&      \\
     0500007896   &   9.296    &   14.0   &14.950 & -1.360 & 100.27596 &    9.41769    &II &	34.70&   0.20	\\
     0500007918   &     I      &    2.7   &15.330 & -1.070 & 100.14539 &    9.90200    &   &	5.200&   0.30	\\
     0500007930   &   6.30     &   19.8   &15.620 & -0.810 & 100.18580 &    9.54061    &   &	60.00&   0.30	\\
     0500007939   &   1.029    &    5.3   &15.430 &  1.200 & 100.24333 &    9.45696    &III&	6.700&      \\
     0500007963   &   2.568    &   28.6   &15.880 & -0.660 & 100.19968 &    9.55087    &   &	      & 	\\
     0500007992   &   2.318    &    2.4   &15.170 & -1.480 & 100.19115 &    9.64566    &III&	4.100&  \\
     0500008003   &   3.19     &    5.0   &15.630 & -1.070 & 100.35450 &    9.60005    &III&	      & 	\\
     0500008007   &   1.805    &    4.4   &15.900 & -0.820 & 100.17437 &    9.69406    &   &	3.100&  \\
     0500008038   &   3.469    &    6.7   &15.660 & -1.150 & 100.10687 &    9.99993    &II &	      &   0.30   \\
     0500008049   &   10.44    &   26.4   &15.430 & -1.420 & 100.32468 &    9.48364    &II &	231.4&   0.40	\\
     0500008061   &     -      &    6.8   &15.650 & -1.230 & 100.32534 &    9.64038    &   &	32.50&   0.30	\\
     0500008064   &     I      &    0.8   &15.930 & -0.970 & 100.22479 &    9.84946    &   &	      & 	\\
     0500008145   &   4.448    &    5.2   &15.620 & -1.590 & 100.16884 &    9.58365    &II &	      & 	\\
     0500008156   &     -      &   20.7   &15.880 & -1.360 & 100.30215 &    9.58578    &   &	      & 	\\
     0500008183   &   7.67     &    3.5   &15.740 & -1.610 & 100.27488 &    9.65395    &II &	7.400&   0.20	\\
     0500008192   &   2.408    &   12.1   &15.760 & -1.620 & 100.30569 &    9.63716    &II &	      & 	\\
     0500008211   &   2.324    &    3.6   &15.770 & -1.700 & 100.26266 &    9.62660    &II &	34.10&   0.50	\\
     0500008213   &   4.364    &    3.9   &15.880 & -1.600 & 100.27111 &    9.82302    &II &	8.300&      \\\hline
\end{tabular}
\begin{flushleft}
$^a$: I=indicates irregular variables, AA=indicates AA Tau-type stars;
$^b$: Short term variability amplitude of the light curves;\\
$^c$: Membership-IR classification (Class II and Class III) of \citet{ssb09}, considering also other memberbership criteria listed in \citet{sbc+08};
$^d$: H$\alpha$ equivalent width from \citet{lbm+04};
$^e$: Masses are from the \citet{sdf00} tracks. Stars were placed in the T$_{eff}$, L$_{bol}$ diagram converting spectral types to T$_{eff}$ and I-band and (in absence of the I magnitude) V-band bolometric corrections, using \citet{kh96}. \\
\end{flushleft}
\end{center}
\end{table*}

\section*{Acknowledgments}
LA and GM acknowledge support from the ASI-INAF agreement I/044/10/0.

\label{lastpage}

\end{document}